\documentclass[aps,prx,reprint,superscriptaddress]{revtex4-1}

\renewcommand\thesection{\arabic{section}}

\usepackage[a4paper,left=2cm,right=2cm,top=3cm,bottom=3cm]{geometry}
\usepackage{soul}
\usepackage{relsize}
\usepackage{lmodern}
\usepackage{slantsc}
\usepackage{graphicx}
\usepackage{amsmath}
\usepackage{amssymb}
\usepackage{amsthm}
\usepackage{amsbsy}
\usepackage{mathrsfs}
\usepackage{varioref}
\usepackage{dsfont}
\usepackage{bm}
\usepackage{color}
\usepackage[usenames,dvipsnames]{xcolor}
\usepackage[colorlinks = false, citecolor=green, linkcolor=red, breaklinks=true, linktocpage=true]{hyperref}
\usepackage[font = footnotesize, labelfont = bf, justification = RaggedRight]{caption}
\usepackage{subfigure}
\usepackage{booktabs} 
\usepackage{array} 
\usepackage{paralist} 
\usepackage{verbatim} 
\edef\restoreparindent{\parindent=\the\parindent\relax}
\usepackage[parfill]{parskip}
\restoreparindent
\usepackage{natbib}

\newcommand{\sub}[1]{_{\!\mathsmaller{\, #1}}}
\newcommand{\eq}[1]{Eq.~\eqref{#1}}
\newcommand{\fig}[1]{Fig.~\ref{#1}}
\newcommand{\sect}[1]{Sec.~\ref{#1}}
\newcommand{\app}[1]{Appendix~(\ref{#1})}

\newcommand{\ts}{\textsuperscript}
\newcommand{\<}{\langle}
\renewcommand{\>}{\rangle}
\newcommand{\ket}[1]{|{#1}\rangle}

\newcommand{\pr}[1]{P[{#1}]}
\newcommand{\prs}[1]{P\sub{\s}[{#1}]}
\renewcommand{\pre}[1]{P\sub{\e}[{#1}]}

\newcommand{\h}{{\mathcal{H}}}

\newcommand{\rr}{{\mathcal{R}}}

\newcommand{\s}{{\mathcal{S}}}
\newcommand{\e}{{\mathcal{E}}}

\renewcommand{\aa}{{\mathcal{A}}}

\newcommand{\nn}{\mathcal{N}}

\newcommand{\one}{\mathds{1}}

\newcommand{\tr}{\mathrm{tr}}

\renewcommand{\ln}[1]{\mathrm{ln} \left( {#1}\right)}
\def\be{\begin{align}}
\def\ee{\end{align}}

\theoremstyle{plain}

\begin{document}

\title{Efficiency of a cyclic quantum heat engine with finite-size baths}

\author{M. Hamed Mohammady}
\affiliation{Department of Physics, Lancaster University, LA1 4YB, United Kingdom}
\affiliation{RCQI, Institute of Physics, Slovak Academy of Sciences, D\'ubravsk\'a cesta 9, Bratislava 84511, Slovakia}


\author{Alessandro Romito}
\affiliation{Department of Physics, Lancaster University, LA1 4YB, United Kingdom}


\begin{abstract}
 In this paper we investigate the relationship between the efficiency of a cyclic quantum heat engine with the Hilbert space dimension of the thermal baths. By means of a general inequality, we show that the Carnot efficiency can  be obtained only when both the hot and cold baths are infinitely large. By further introducing a specific model where the baths are constituted of ensembles of finite-dimensional particles, we further demonstrate the relationship between the engine's power and efficiency, with the dimension of the working substance and the bath particles.   

\end{abstract}

\maketitle

\section{Introduction}

An understanding of the fundamental limitations of heat engines was the initial impetus for the development of thermodynamics.  Indeed, in one of its earliest formulations due to Carnot \cite{Carnot}, the second law of thermodynamics sets an upper bound to the efficiency of a cyclic heat engine, in which  a working substance $\s$ extracts work by  transferring heat from a hot bath, of temperature $T_h$, to a cold bath of temperature $T_c < T_h$. The efficiency of average work, defined as the ratio of average work output to the average heat extracted from the hot bath, is limited by the Carnot efficiency $\eta\sub{C} := 1 - T_c/T_h$.

Recent advances in the technological ability of engineering nano-scale devices demands for extending such thermodynamic principles to small scales where thermal and quantum fluctuations dominate. 
Extensions of the second law of thermodynamics in the form of fluctuation theorems have successfully accounted for thermal fluctuations  \cite{doi:10.1146/annurev-conmatphys-062910-140506, Seifert_2012, doi:10.1146/annurev-conmatphys-030212-184240, PhysRevLett.78.2690}, allowing for the fluctuating behaviour of small-scale heat engines to be addressed \cite{Verley2014, Polettini2015,Martinez2016,Saha2017}. 
The extension of these results to account for quantum fluctuations has led to a growing interest in the thermodynamic properties of quantum systems \cite{Anders-thermo-review,Goold-thermo-review,Millen-thermo-review}, and the development of stochastic quantum thermodynamics so as to extend fluctuation theorems into the quantum domain \cite{PhysRevA.86.044302,Aberg2016,Abdelkhalek2016a,Alexia-measurement-thermodynamics, Naghiloo2018, Elouard2018}. A number of works have addressed the role of quantum mechanical phenomena such as coherence, entanglement, and measurement-induced back-action, on the thermodynamic properties of systems \cite{Sagawa2009b, thermo-individual-quantum,  Lostaglio2015, Guryanova2015,Karen-extractable-work-correlations}, while the properties of quantum heat engines in particular have  received much attention \cite{Scully2011, Uzdin2015,Yi2017,Elouard2017a,Campisi2018, Ghosh2017}.

In most quantum mechanical treatments of heat engines, only the working substance is assumed to be a small, microscopic system, while the thermal baths are assumed to be infinitely large.  It is known that the size of the thermal bath introduces correction terms in the second law of thermodynamics \cite{Richens2018}, and imposes limitations on thermodynamic operations such as cooling  \cite{Allahverdyan2011, Reeb_2014, Scharlau2018}.  Similarly, several works have analysed the finite-size effects of the thermal bath on the performance of heat engines  \cite{Izumida2014,Tajima2017, Reid_2017, Ito2018, Pozas_Kerstjens_2018}. However, these studies have understood the size of the thermal bath to be the number of particles that constitute it, the volume of the bath, or the heat capacity of the bath. A description of how the Hilbert space dimension of the bath affects the performance of the engine, in a manner similar to how the dimension of the bath affects Landauer's principle in Ref. \cite{Reeb_2014}, remains an open problem. Therefore, in this paper we attempt to close this gap by investigating how the Hilbert space dimension of the hot and cold baths affects the performance of a  cyclic quantum heat engine in terms of its efficiency. 

Specifically, in \sect{section:general} we describe a cyclic quantum heat engine in general terms, and quantify how the dimension of the thermal baths limits the efficiency of average work for such an engine in terms of an inequality (\eq{eq:Efficiency-finite-size-General}). This shows that the efficiency of average work can  approach the Carnot efficiency only when both the hot and cold baths have an infinite-dimensional Hilbert space. Subsequently, in \sect{section:number-conserving engine} we introduce a specific model for a cyclic quantum heat engine, inspired by the collision model approach to open system dynamics \cite{Scarani2002ThermalizingQM,Ziman-collision,Pezzutto_2016}. Here, the hot and cold baths are considered as ensembles of particles with equally spaced  energy levels, and the collision  between the working substance and each of these particles is described by a joint unitary evolution that  conserves the total excitation number. We then proceed to show how the efficiency of average work will approach the Carnot limit when the number of ensembles goes to infinity, corresponding with a smooth change in the particle energy gaps. This will result in the dimension of the baths to approach infinity.  In \sect{Dimension of thermal baths and the engine's performance} we quantitatively explore the relationship between the dimension of the working substance $\s$ and that of the baths, with the power and efficiency of the engine. We consider two classes of collision interactions: (i) a swap operation, and (ii) a unitary generated by a Jaynes-Cummings Hamiltonian.  Finally, in \sect{Achieving the stochastic Carnot efficiency with finite-dimensional baths} we analyse the stochastic efficiency of this engine. We show that the efficiency of the most likely trajectory per cycle will approach the Carnot efficiency from below as the dimension of the cold bath becomes infinitely large; the size of the hot bath does not affect this. Meanwhile, the most likely stochastic efficiency per cycle (distinct from the efficiency of the most likely trajectory, since multiple trajectories can have the same efficiency) can be the Carnot efficiency when both hot and cold baths are small, finite-dimensional systems.

\section{General constraints on work extraction and efficiency} \label{section:general}

The system we are interested in is a cyclic engine operating between two thermal baths at different temperatures.
The  engine is a working substance $\s$ with Hilbert space $\h\sub{\s}$ and the two thermal baths, $\rr$ and $\e$, have Hilbert spaces $\h\sub{\rr}$ and $\h\sub{\e}$, respectively. The compound system is described by the time-dependent Hamiltonian $H(t) := H\sub{\s} + H\sub{\rr} + H\sub{\e} + V(t)$, which changes smoothly with $t$. The time-dependence of the Hamiltonian is to be understood as effecting an exchange of work with an external work storage device which we do not explicitly include within the quantum description. At time $t=0$ the working substance is decoupled from the thermal baths, $V(0)=0$, and the compound system is in the product state $\rho(0) := \rho\sub{\s}(0)\otimes \rho\sub{\rr}(0)\otimes \rho\sub{\e}(0)$, such that $\e$ and $\rr$ are at thermal equilibrium with respect to their bare Hamiltonians, i.e. for $X \in \{\rr,\e\}$, $\rho\sub{X}(0):= e^{-\beta\sub{X} H\sub{X}}/\tr[e^{-\beta\sub{X} H\sub{X}}]$ with inverse temperature $\beta\sub{X}$. We assume that $\e$ is warmer than $\rr$, i.e.  $\beta\sub{\e} < \beta\sub{\rr}$. The time dependence of the total Hamiltonian lets the compound system evolve as $\rho(t) := U(t) \rho(0) U^\dagger(t)$ with the unitary time-evolution operator $U(t):= \underleftarrow{\mathcal{T}} e^{-i \int_0^t d \tau H(\tau)}$. Here $\underleftarrow{\mathcal{T}}$ denotes the time ordering operator. The reduced state of each  subsystem $X \in \{\s,\rr,\e\}$ at time $t$ is thus given by $\rho\sub{X}(t) := \tr\sub{\bar X}[\rho(t)]$, where $\tr\sub{\bar X}$ denotes the partial trace over all systems other than $X$.

The average work extracted from the compound system, during the time interval $[0,T]$, can be calculated \cite{Alicki_1979} to be 
\begin{align}\label{eq:average-work-general}
\<W\> &= \int_T^0 dt \, \, \tr\left[\frac{d H(t)}{dt}  \rho(t) \right] , \nonumber \\
& = \int_T^0 dt \, \, \frac{d}{dt}\tr[ H(t)\rho(t) ] - \tr\left[H(t) \frac{d \rho(t)}{dt}\right], \nonumber \\
&=  \tr[H(0) \rho(0)] - \tr[H(T) \rho(T)].
\end{align}
 Here, the last step is obtained by noting that the compound system evolves unitarily, and hence $\frac{d \rho(t)}{dt} = i [H(t), \rho(t)]$. Therefore,  by the cyclicity property of the trace we have  $\tr\left[H(t) \frac{d \rho(t)}{dt}\right] = i \tr[H(t) (H(t) \rho(t) - \rho(t) H(t))] = 0$. Recall that $\rho(T) := U(T) \rho(0) U^\dagger(T)$, and so \eq{eq:average-work-general} shows that the average work extracted is simply the change in internal energy of the compound system $\s+\rr+\e$ as it unitarily evolves by $U(T)$.

Our heat engine will be cyclic (with cycle time $T$) if it satisfies two conditions:  (i) $H(T)$ = $H(0) = H\sub{\s} + H\sub{\rr} + H\sub{\e}$ and (ii) $\rho\sub{\s}(T) = \rho\sub{\s}(0)$. Condition (i) means that the interaction between the subsystems is switched on at time $t=0^+$ and switched off at time  $t=T^-$. Since the Hamiltonian $H(0) = H(T)$ is additive,   \eq{eq:average-work-general} will reduce to
\begin{align}\label{eq:average-work-general-same-Hamiltonian}
\<W\>  &= \tr[H(0) (\rho(0) -\rho(T))], \nonumber \\
& = \sum_{X \in \{\s,\rr,\e\}}\tr[H\sub{X} (\rho\sub{X}(0) - \rho\sub{X}(T))], \nonumber \\
& = \<\Delta E\sub{\s}\> +  \<Q\sub{\e}\> - \<Q\sub{\rr}\>.
\end{align}
Here we identify $ \<\Delta E\sub{\s}\>$ as the average decrease in internal energy of $\s$, while $\<Q\sub{\e}\>$ ($\<Q\sub{\rr}\> $) is the average heat absorbed from (by) the thermal bath $\e$ ($\rr$). Condition (ii), meanwhile, implies that the internal energy of the working substance is the same at the start and end of the cycle, i.e. $\<\Delta E\sub{\s}\> =0$. Therefore, we are left with  
\begin{align}\label{eq:average-work-cyclic-general}
\<W\> &= \<Q\sub{\e}\> - \<Q\sub{\rr}\>.
\end{align}

Since the initial states of the baths are given by Gibbs states, we may express the heat terms  as 
\begin{align}\label{eq:ent-relent-bath}
\<Q\sub{\e}\> &= \frac{1}{\beta\sub{\e}}\left(\Delta S\sub{\e} - D[\rho\sub{\e}(T) \| \rho\sub{\e}(0)]\right), \nonumber \\
\<Q\sub{\rr}\> &= \frac{1}{\beta\sub{\rr}}\left( D[\rho\sub{\rr}(T) \| \rho\sub{\rr}(0)] - \Delta S\sub{\rr} \right).
\end{align}
Here $\Delta S\sub{X} := S(\rho\sub{X}(0)) - S(\rho\sub{X}(T))$ is the decrease in von Neumann entropy, $S(\rho):= -\tr[\rho \, \ln{\rho}]$, of  system $X \in \{\rr,\e\}$, and $D[\rho \| \sigma]:= \tr[\rho(\ln{\rho} - \ln{\sigma})]$ is the entropy of $\rho$ relative to $\sigma$ \cite{Sagawa2012a}.
As the relative entropy is non-negative, it follows from \eq{eq:average-work-cyclic-general} and \eq{eq:ent-relent-bath} that 
\begin{align}\label{eq:lavoro}
\<W\> &\leqslant \frac{1}{\beta\sub{\e}}\Delta S\sub{\e}  +  \frac{1}{\beta\sub{\rr}}\Delta S\sub{\rr} . 
\end{align}
Moreover, due to the sub-additivity of the von Neumann entropy, and its preservation under unitary evolution \cite{DenesPetz2008}, it follows that $\Delta S\sub{\e} + \Delta S\sub{\rr} := -S_\mathrm{irr} \leqslant 0$, where $S_\mathrm{irr}$ denotes the irreversible entropy production.  
Consequently, by \eq{eq:lavoro} and the restriction on the irreversible entropy production,   the average work extraction will be positive only if
$-\Delta S\sub{\rr} \geqslant \Delta S\sub{\e} > -\frac{\beta\sub{\e}}{\beta\sub{\rr}}\Delta S\sub{\rr}$. Since $\beta\sub{\e} < \beta\sub{\rr}$ by construction, therefore, this inequality will be satisfied only if $\Delta S\sub{\e} >0$ and $\Delta S\sub{\rr} <0$. Finally,  these inequalities in conjunction with  \eq{eq:average-work-cyclic-general} and \eq{eq:ent-relent-bath} indicate  that a cyclic heat engine produces positive work on average only if 
\begin{align}\label{eq:heat-inequality-positive-work}
\<Q\sub{\e}\>  > \<Q\sub{\rr}\>  >0.
\end{align}

\subsection{How the bath dimension restricts the efficiency}

The efficiency of converting heat from the hot bath into work is given by \eq{eq:average-work-cyclic-general} as
\begin{align}\label{eq:efficiency}
\eta: = \frac{\<W\>}{ \<Q\sub{\e}\>} = 1 - \frac{\<Q\sub{\rr}\>}{\<Q\sub{\e}\>}.
\end{align}
Given \eq{eq:heat-inequality-positive-work}, therefore, it is impossible for a cyclic heat engine to operate at unit efficiency, i.e., to fully convert heat from the hot bath into work; there will always be some residual heat that is transferred to the cold bath. This is in accordance with the second law of thermodynamics. 

In order to explore how the dimension of the heat baths affect the efficiency of a cyclic heat engine,  we use the results of Ref. \cite{Reeb-Wolf-RelEnt}, which showed that $D[\rho\| \sigma] \geqslant \frac{\Delta S^2}{3 \mathrm{ln}^2(\dim(\h))}$, where $\Delta S := S(\sigma) - S(\rho)$, and $\dim(\h)$ is the dimension of the Hilbert space on which $\rho$ and $\sigma$ act. Therefore, by \eq{eq:ent-relent-bath},  and the definition of the irreversible entropy production $S_\mathrm{irr}$, we show that the efficiency of a cyclic heat engine producing positive work on average obeys the inequality
\begin{align}\label{eq:Efficiency-finite-size-General}
\eta &= 1 - \frac{\beta\sub{\e}}{\beta\sub{\rr}}\left(\frac{S_\mathrm{irr} + \Delta S \sub{\e} + D[\rho\sub{\rr}(T)\|\rho\sub{\rr}(0) ] }{ \Delta S\sub{\e} - D[\rho\sub{\e}(T)\|\rho\sub{\e}(0) ] }\right), \nonumber \\
& \leqslant 1 - \frac{\beta\sub{\e}}{\beta\sub{\rr}}\left(\frac{  S_\mathrm{irr} + \Delta S \sub{\e} +  \frac{\Delta S\sub{\rr}^2 }{3  \mathrm{ln}^2(\dim(\h\sub{\rr}))}  }{ \Delta S\sub{\e} - \frac{\Delta S\sub{\e}^2}{3  \mathrm{ln}^2(\dim(\h\sub{\e}))}}   \right), \nonumber \\
& \leqslant 1 - \frac{\beta\sub{\e}}{\beta\sub{\rr}}=: \eta\sub{C}, 
\end{align}
where $\eta\sub{C}$ denotes the familiar Carnot efficiency. The requirement that the average work be positive demands that both the numerator and denominator in the fraction appearing on the first line of \eq{eq:Efficiency-finite-size-General} must be positive. Consequently, since $\Delta S\sub{\e}>0$ and $S_\mathrm{irr}>0$,  the inequality in the second line follows from taking the lower bounds of the relative entropy terms, which effectively determines the maximum value of $\<Q\sub{\e}\> >0$ and the minimum value of $\<Q\sub{\rr}\> > 0$ given the corresponding entropy changes.  The equality in the third line is achieved by taking the limits $\dim(\h\sub{X}) \to \infty$ and $S_\mathrm{irr} \to 0$. 

The inequality in the second line of \eq{eq:Efficiency-finite-size-General} quantifies how finite bath dimensions limits the efficiency of average work. In order to better understand this relationship,  we shall turn to a specific model which we introduce in the next section. 

\section{A  cyclic heat engine with finite-sized baths}\label{section:number-conserving engine}

 \begin{figure}[!htb]
\includegraphics[width = 0.45\textwidth]{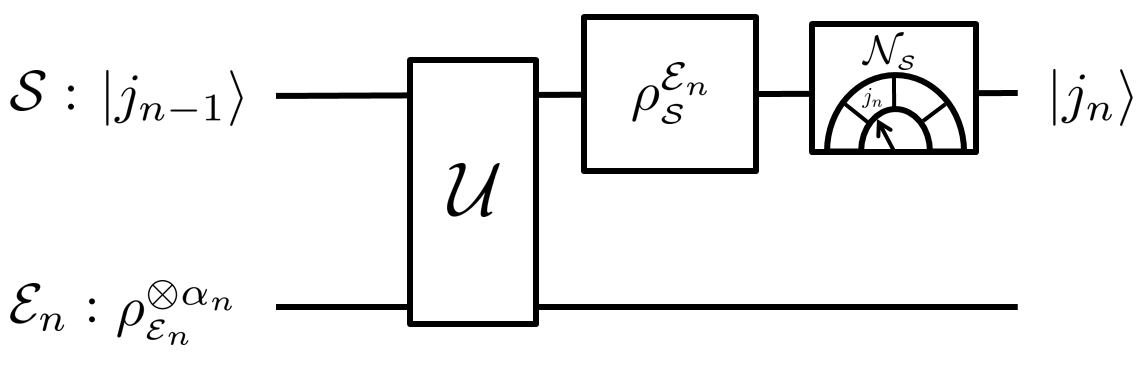}
\caption{ Here we sketch a single step of the cyclic heat engine. Initially, the working substance $\s$ is prepared in an eigenstate of its number operator, denoted $\ket{j_{n-1}}$. Subsequently, $\s$ sequentially interacts with $\alpha\sub{n}$ particles from the ensemble $\e\sub{n}$, with the number conserving unitary $U$. Here, $\mathcal{U}$ denotes the consecutive applications of $U$. For a sufficiently large $\alpha\sub{n}$, the resulting state of $\s$ will be approximated by the pseudo-thermal state $\rho\sub{\s}^{\e\sub{n}}$ defined by \eq{eq:pseudo-thermal-state}. Finally, $\s$ is projectively measured with respect to its number operator, being prepared in the state $\ket{j_n}$. Due to the number conservation of $U$, it follows that the heat absorbed from the ensemble $\e\sub{n}$ is $\omega\sub{n}(j_n - j_{n-1})$.  }\label{fig:Setup}
\end{figure}  

In order to quantitatively study how the finite dimensions of $\h\sub{\e}$ and $\h\sub{\rr}$ affect a cyclic heat engine,  we turn to a simple  example where the hot and cold baths are described as ensembles of ``particles'' with a finite-size Hilbert space,  which interact sequentially with the working substance $\s$ akin to collision models \cite{Scarani2002ThermalizingQM,Ziman-collision,Pezzutto_2016}. Moreover, we shall restrict the collision model to one where each bath particle has equally spaced energy levels (such as a truncated harmonic oscillator), and the interaction between $\s$ and the bath particles conserves the total excitation number. While not essential, we shall show that such a restriction leads to interesting consequences such as the ability of defining stochastic work, heat, and efficiency by only performing projective  measurements on the  working substance, and constructing an engine that approaches the Carnot efficiency arbitrarily well with minimal control of the system-bath interactions.  

The model  we consider is sketched in \fig{fig:Setup}. The working substance $\s$ has a Hilbert space with dimension $d\sub{\s}$, while  the hot and cold baths $\e$ and $\rr$ are comprised of $N$ and $M$ ensembles of identical systems (particles),  labeled as $\{\e_n: n \in \{1, \dots, N\} \}$ and $\{\rr_m: m \in \{1,\dots,M\}\}$ respectively. The  particles of the ensemble $\e_n$ ($\rr_m$) have the same Hilbert space dimension $d\sub{\e}$ ($d\sub{\rr}$), but generically differ by their Hamiltonians 
\begin{align}
H\sub{\e_n} &:=  \sum_{k=0}^{d\sub{\e} -1}  E^{\e_n}_k \pr{k}, \nonumber \\
H\sub{\rr_m} &:=  \sum_{k=0}^{d\sub{\rr} -1} E^{\rr_m}_k \pr{k}.
\end{align}
Here, $\pr{k} \equiv |k\>\<k|$ is a projection on vector $\ket{k}$ denoting $k$ quanta of excitation. We further assume that for all $k$,  $E^{\e_n}_{k+1} - E^{\e_n}_k = \omega\sub{n}>0$ and $E^{\rr_m}_{k+1} - E^{\rr_m}_k = \Omega\sub{m}>0$. Therefore,  we may equivalently express the Hamiltonians as  $H\sub{\e_n}=\omega\sub{n} \nn\sub{\e}$ and $H\sub{\rr_m}=\Omega\sub{m} \nn\sub{\rr}$, which are proportional to their number operators 
\begin{align}\label{eq:Number-operator}
\nn\sub{X} &:= \sum_{k=0}^{d\sub{X} -1} k  \pr{k}. 
\end{align}

Before interacting with $\s$, each particle in ensemble $\e\sub{n}$  is in the thermal state $\rho\sub{\e\sub{n}}:= e^{- \beta\sub{\e} \omega\sub{n}  \nn\sub{\e}}/\tr[e^{- \beta\sub{\e} \omega\sub{n} \nn\sub{\e}}]$.
The collision between $\s$ and these particles is described by a unitary evolution $U$ which conserves the total number, i.e.,  commutes with $\nn\sub{\s} + \nn\sub{\e}$, and the corresponding quantum channel on $\s$ is denoted as 
\begin{align}
\Lambda_{\e\sub{n}}: \rho\sub{\s} \mapsto \tr\sub{\bar \s}[U(\rho\sub{\s} \otimes \rho\sub{\e\sub{n}})U^\dagger].
\end{align}
 As shown in \app{app:stationary state} the stationary state of $\s$, given the quantum channel  $\Lambda_{\e\sub{n}}$, is
\begin{align}\label{eq:pseudo-thermal-state}
\rho\sub{\s}^{\e\sub{n}}:= \frac{e^{-\beta\sub{\e} \omega\sub{n} \nn\sub{\s}}}{\tr[e^{-\beta\sub{\e} \omega\sub{n} \nn\sub{\s}}]}.
\end{align}
The interaction with the cold bath ensembles $\rr\sub{m}$ is defined analogously.  Note that,  in general, the Hamiltonian of $\s$ is arbitrary and need not commute with its number operator $\nn\sub{\s}$. Consequently, while $\rho\sub{\s}^{\e\sub{n}}$ and $\rho\sub{\s}^{\rr\sub{m}}$ need not be thermal states of $\s$, for convenience we call them ``pseudo-thermal'' given that they can be written as a Gibbs state with respect to the number operator \cite{Mohammady2017a}. 
Moreover, the stationary state is approximated with arbitrary precision by a finite consecutive application of   $\Lambda_{\e\sub{n}}$ (see \app{app:stationary state}). 
Therefore, we say that $\s$ has $\epsilon$-pseudo-thermalized to the state $\rho\sub{\s}^{\e\sub{n}}$ ($\rho\sub{\s}^{\rr\sub{m}}$), if its trace distance to this state is less than $\epsilon$.

Now we may define each cycle of the engine as a sequence of pseud-thermalizations with the bath ensembles. The initial state of $\s$ at the start of the cycle is $\rho\sub{\s}^{\rr\sub{M}}$, which can be obtained by letting $\s$ pseudo-thermalize by interacting with the cold bath ensemble $\rr\sub{M}$. The cycle then consists of three steps:
\begin{enumerate}[(i)]
	\item   Projectively measure $\s$ with respect to the number operator $\nn\sub{\s}$, which prepares the system in the pure state $\ket{j_0}$. 
	\item For $n$ running from $1$ to $N$: let the system $\epsilon$-pseudo-thermalize to $\rho\sub{\s}^{\e\sub{n}}$ by $\alpha\sub{n}$ number conserving interactions with particles from the hot bath ensemble $\e\sub{n}$, and then projectively measure the number of $\s$, preparing it in the pure state $\ket{j_n}$. 
	\item  For $m$ running from $1$ to $M$:  let the system $\epsilon$-pseudo-thermalize to $\rho\sub{\s}^{\rr\sub{m}}$ by $\alpha\sub{m}$ number conserving interactions with particles from the cold bath ensemble $\rr\sub{m}$, and then projectively measure the number of $\s$, preparing it in the pure state $\ket{k_m}$ (note that $\ket{j_N}=\ket{k_0}$).  
\end{enumerate}
As the trace distance between the initial and final state of $\s$ is smaller than $\epsilon$, we refer to this engine as  $\epsilon$-cyclic. Furthermore, the results of the projective measurements of $\s$ by the number operator constitutes a trajectory of the engine, labeled as $\gamma := (j_0,\dots,j_N \equiv k_0,k_1,\dots,k_M)$. Since the state of $\s$ at the start of the cycle commutes with $\nn\sub{\s}$, the dynamics ensures that its state will always commute with $\nn\sub{\s}$ throughout the cycle  (see \app{app:stationary state}). 
Consequently, the projective measurements of $\s$ by the number operator will not disturb the state of the system, and the average evolution of this system can indeed be seen as a probabilistic evolution along the trajectories $\gamma$.

In the simplest case of $N=M=1$, steps (ii) and (iii) of the cycle involve interactions with identical particles, namely, $ \rho\sub{\e\sub{1}}$ and $ \rho\sub{\rr\sub{1}}$, respectively. As such, the cycle can be thought of as being a collision model analogue  to the classical setup where $\s$ is first brought to thermal equilibrium with the hot bath $\e$, and then brought to thermal equilibrium with the cold bath $\rr$. However, when $N=M=1$, the efficiency will necessarily be sub-Carnot, and the only way to approach  the Carnot efficiency is to produce a vanishingly small amount of work.   However, we shall see that by increasing $N$ and $M$, resulting in small increments in $\omega\sub{n}$ and $\Omega\sub{m}$, the engine will approach the quasistatic limit, allowing for the efficiency to approach  the Carnot limit while still producing a finite amount of work.

\subsection{Work, heat, and efficiency of  the engine}

The fact that the bath particle Hamiltonians are proportional to their number, and that the interaction between $\s$ and these particles conserves the total number of excitations, allows us to evaluate heat from the measurements on $\s$ alone -- no measurements of the bath particles are required. Indeed,  when the working substance's number increases as $j_n - j_{n-1}$, the hot bath ensemble $\e\sub{n}$ loses $\omega\sub{n}(j_n - j_{n-1})$ quanta of energy, and when the working substance's number increases as $k_m - k_{m-1}$, the cold bath ensemble $\rr\sub{m}$ absorbs  $\Omega\sub{m}(  k_{m-1} - k_m)$ quanta of energy. 

The heat values for each trajectory are thus given as 
\begin{align}\label{eq:heat-traj}
Q\sub{\e}(\gamma) &= \sum_{n=1}^N \omega\sub{n}(j_n - j_{n-1}), \nonumber \\
Q\sub{\rr}(\gamma) & = \sum_{m=1}^M \Omega\sub{m}(k_{m-1} - k_m) .
\end{align}
The decrease in internal energy of the system, meanwhile, is simply $\Delta E(\gamma) =\<j_0|H\sub{\s}|j_0\> - \<k_M| H\sub{\s} |k_M\>$. Therefore, the work for each trajectory will be given by the first law of thermodynamics as
\begin{align}\label{eq:work-traj}
W(\gamma) &= \Delta E(\gamma) + Q\sub{\e}(\gamma) - Q\sub{\rr}(\gamma).
\end{align}
For the subset of trajectories $\gamma$ such that $W(\gamma) \ne 0$ and $Q\sub{\e}(\gamma) \ne 0$, we may define the stochastic efficiency as
\begin{align}\label{eq:efficiency-traj}
\eta(\gamma) &:= \frac{W(\gamma)}{Q\sub{\e}(\gamma)} , \nonumber \\
& = 1 +  \frac{\sum_{m=1}^M \Omega\sub{m}(k_m - k_{m-1}) + \Delta E(\gamma)}{\sum_{n=1}^N \omega\sub{n}(j_n - j_{n-1})}.
\end{align}
Finally, in the limit as $\epsilon \to 0$, wherein the engine is completely cyclic,   the probability of each trajectory  $\gamma$ is given by
\begin{align}\label{eq:prob-traj}
p(\gamma) &= \<j_0|\rho\sub{\s}^{\rr\sub{M}}|j_0\>   \prod_{n=1}^N \<j_n|\rho\sub{\s}^{\e\sub{n}}|j_n\> \nonumber \\
& \qquad \times \prod_{m=1}^M \<k_m|\rho\sub{\s}^{\rr\sub{m}}|k_m\>.
\end{align}

From Eqs. (\ref{eq:prob-traj},\ref{eq:work-traj}, \ref{eq:heat-traj}) we can determine the average performance of the engine.  First, we note that since the engine is cyclic, the average decrease in the internal energy of the working substance  is 
\begin{equation}
\<\Delta E\>:= \sum_\gamma p(\gamma)\Delta E(\gamma) = \tr[H\sub{\s} (\rho\sub{\s} ^{\rr_M} - \rho\sub{\s} ^{\rr_M})]  = 0.
\end{equation}
As such, the average work is given by \eq{eq:work-traj} as    $\<W\> = \<Q\sub{\e}\> - \<Q\sub{\rr}\>$, in concordance  with \eq{eq:average-work-cyclic-general}. Here, the average heat absorbed from the hot bath is
\begin{align}\label{eq:avg-heat-hot-oscillator}
\<Q\sub{\e}\> &= \omega\sub{1} \tr[\nn\sub{\s}(\rho\sub{\s}^{\e\sub{1}} - \rho\sub{\s}^{\rr\sub{M}})] \nonumber \\
& \qquad + \sum_{n=2}^N \omega\sub{n} \tr[\nn\sub{\s}(\rho\sub{\s}^{\e\sub{n}} - \rho\sub{\s}^{\e\sub{n-1}})], \nonumber \\
& = \frac{1}{\beta\sub{\e}}\left(\Delta S - D\sub{\e}\right),
\end{align}
while the average heat transferred to the cold bath is
\begin{align}\label{eq:avg-heat-cold-oscillator}
\<Q\sub{\rr}\> &= \Omega\sub{1} \tr[\nn\sub{\s}(\rho\sub{\s}^{\e\sub{N}} - \rho\sub{\s}^{\rr\sub{1}})] \nonumber \\
& \qquad + \sum_{m=2}^M \Omega\sub{m} \tr[\nn\sub{\s}(\rho\sub{\s}^{\rr\sub{m-1}} - \rho\sub{\s}^{\rr\sub{m}})], \nonumber \\
& = \frac{1}{\beta\sub{\rr}}\left(\Delta S + D\sub{\rr}\right),
\end{align}
where  $\Delta S := S(\rho\sub{\s}^{\e\sub{N}}) - S(\rho\sub{\s}^{\rr\sub{M}})$, $D\sub{\e} := D[\rho\sub{\s}^{\rr\sub{M}}\|\rho\sub{\s}^{\e\sub{1}}] + \sum_{n=2}^N D[\rho\sub{\s}^{\e\sub{n-1}}\|\rho\sub{\s}^{\e\sub{n}}] $, and $D\sub{\rr} := D[\rho\sub{\s}^{\e\sub{N}}\|\rho\sub{\s}^{\rr\sub{1}}] + \sum_{m=2}^M D[\rho\sub{\s}^{\rr\sub{m-1}}\|\rho\sub{\s}^{\rr\sub{m}}]$. Therefore,  we may express the average work and efficiency as 
\begin{align}\label{eq:avg-work-oscillator}
\<W\> =  \left(\frac{1}{\beta\sub{\e}} - \frac{1}{\beta\sub{\rr}}\right) \Delta S - \frac{D\sub{\e}}{\beta\sub{\e}} -  \frac{D\sub{\rr}}{\beta\sub{\rr}},
\end{align}
and 
\begin{align}\label{eq:avg-efficiency-oscillator}
\eta = 1 - \frac{\beta\sub{\e}}{\beta\sub{\rr}}\left(\frac{\Delta S + D\sub{\rr}}{\Delta S - D\sub{\e}} \right).
\end{align}
Note the similarity between \eq{eq:avg-efficiency-oscillator} and \eq{eq:Efficiency-finite-size-General} except that, in the former, the entropic quantities  pertain to $\s$ and not  the thermal baths. Given the positivity of the relative entropy terms $D\sub{\e}$ and $D\sub{\rr}$, we immediately arrive at the inequality $\<W\> \leqslant \left(\frac{1}{\beta\sub{\e}} - \frac{1}{\beta\sub{\rr}}\right) \Delta S$, implying that positive work extraction is  possible only when $\Delta S >0$ which, in turn, requires that  $\beta\sub{\e} \omega\sub{N} < \beta\sub{\rr} \Omega\sub{M}$. Additionally, given a fixed value of $\Delta S$, both the average work and the efficiency are maximised by taking the limits $D\sub{\e} \to 0$ and $D\sub{\rr} \to 0$. In this limit the efficiency approaches the Carnot value.

\subsection{Limiting case 1: baths comprised of single ensembles}

\begin{figure}[!htb]
\includegraphics[width = 0.45\textwidth]{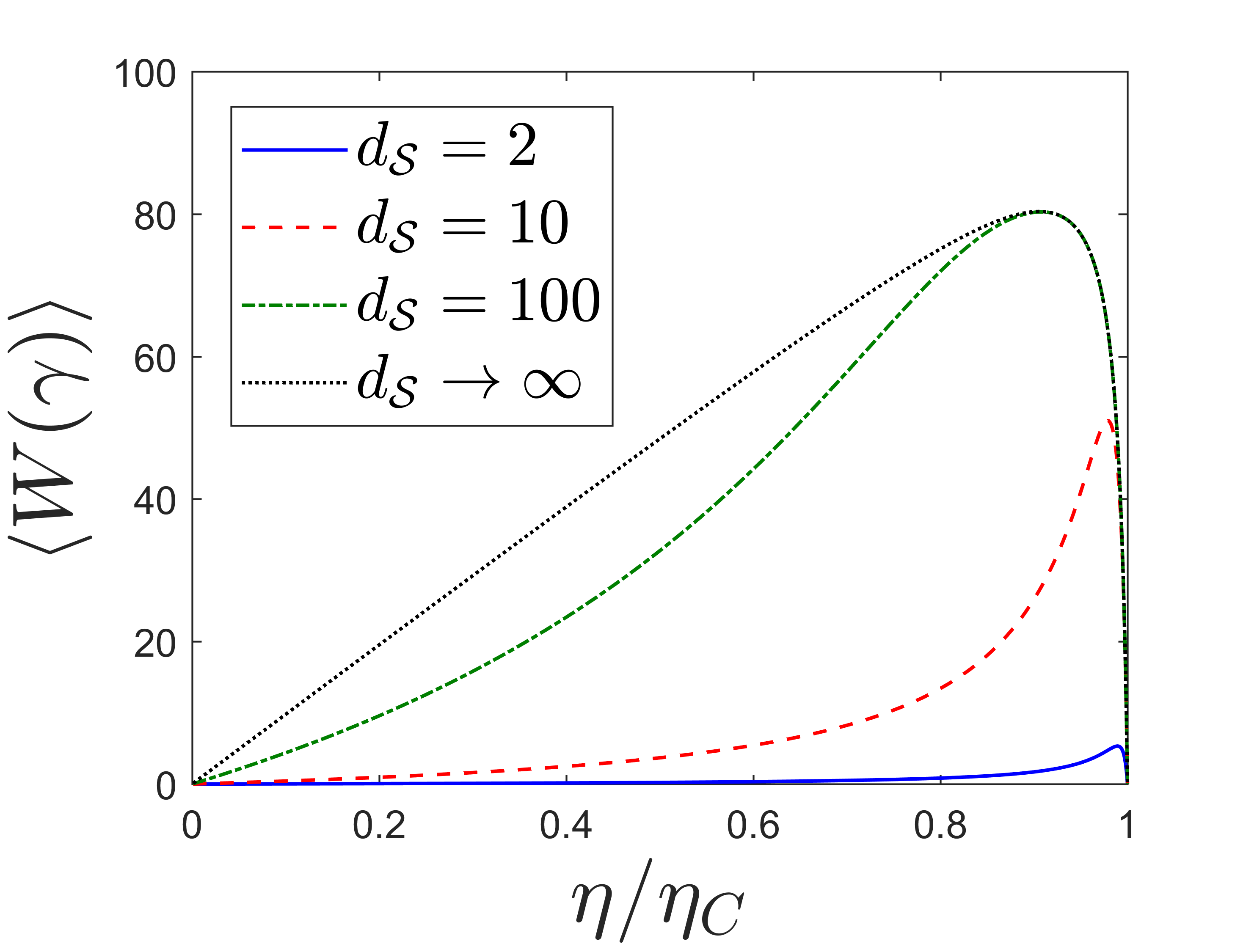}
\caption{Relationship between average work and efficiency when the hot and cold baths are comprised of single ensembles, with  frequencies $\omega\sub{1}$ and $\Omega\sub{1}$ respectively. Here we fix $\beta\sub{\rr} = 1$, $\Omega\sub{1} = 1$, $\beta\sub{\e} = 10^{-2}$, and only vary the value of $\omega\sub{1}$. The maximum work increases with the dimension of the working substance, $d\sub{\s}$, whereas the efficiency at maximum work decreases with $d\sub{\s}$.  }\label{fig:Work_Eff_single_ensemble}
\end{figure}

Let us consider the limiting case where the hot and cold baths are comprised of single ensembles, i.e., when  $N=M=1$.  
In this special case, the average work and efficiency obtain the simple expressions 
\begin{align}\label{eq:work-efficiency-single-ensemble}
\<W\> &= (\omega\sub{1} - \Omega\sub{1})\tr[\nn\sub{\s}(\rho\sub{\s}^{\e\sub{1}} - \rho\sub{\s}^{\rr\sub{1}} )],\nonumber \\
\eta &= 1 - \frac{\Omega\sub{1}}{\omega\sub{1}}.
\end{align}
The expression for the efficiency obtained above is similar to that of \cite{Yi2017}, except that the frequencies here pertain to the thermal baths, and not the working substance. \eq{eq:work-efficiency-single-ensemble} shows that the efficiency approaches the Carnot value when $\omega\sub{1} = (\beta\sub{\rr}/\beta\sub{\e})\Omega\sub{1}$, implying that $\rho\sub{\s}^{\e\sub{1}} = \rho\sub{\s}^{\rr\sub{1}}$. This results in a trivial engine with  $\Delta S = 0$ and hence zero average work extraction. We may also make this observation by directly appealing to \eq{eq:avg-efficiency-oscillator}, where the Carnot efficiency is achieved only when $D\sub{\e} =  D[\rho\sub{\s}^{\rr\sub{1}}\|\rho\sub{\s}^{\e\sub{1}}] = 0 $ and $D\sub{\rr} = D[\rho\sub{\s}^{\e\sub{1}}\|\rho\sub{\s}^{\rr\sub{1}}]=0$. This can  be achieved only when  $\rho\sub{\s}^{\e\sub{1}} = \rho\sub{\s}^{\rr\sub{1}}$, resulting in $\Delta S = 0$.  

 In \fig{fig:Work_Eff_single_ensemble} we report the relationship between the efficiency  and the average work for different   dimensions of the working substance. Here, we  fix the parameters $\beta\sub{\rr} = 1$, $\Omega\sub{1} = 1$, $\beta\sub{\e} = 10^{-2}$, and  vary only the value of $\omega\sub{1}$. When $\omega\sub{1} = \Omega\sub{1}$, both the work and efficiency vanish. Conversely, when $\omega\sub{1} = (\beta\sub{\rr}/\beta\sub{\e})\Omega\sub{1}$, the efficiency approaches the Carnot value, but the work vanishes, as we discussed previously. The work is maximised when $\omega\sub{1}$ takes a value between these extreme ranges.  Meanwhile, the work obtained for a given efficiency increases with the dimension of the working substance, converging as 
\begin{align}
\lim_{d\sub{\s} \to \infty}\<W\> = \frac{(\omega\sub{1} - \Omega\sub{1})(e^{\beta\sub{\rr} \Omega\sub{1}} - e^{\beta\sub{\e} \omega\sub{1}})}{(e^{\beta\sub{\rr} \Omega\sub{1}}-1) (e^{\beta\sub{\e} \omega\sub{1}} -1)}.
\end{align}
However, the efficiency at maximum work decreases with the dimension of the working substance.

\subsection{Limiting case 2: baths comprised of infinitely many ensembles}
\begin{figure}[!htb]
\includegraphics[width = 0.45\textwidth]{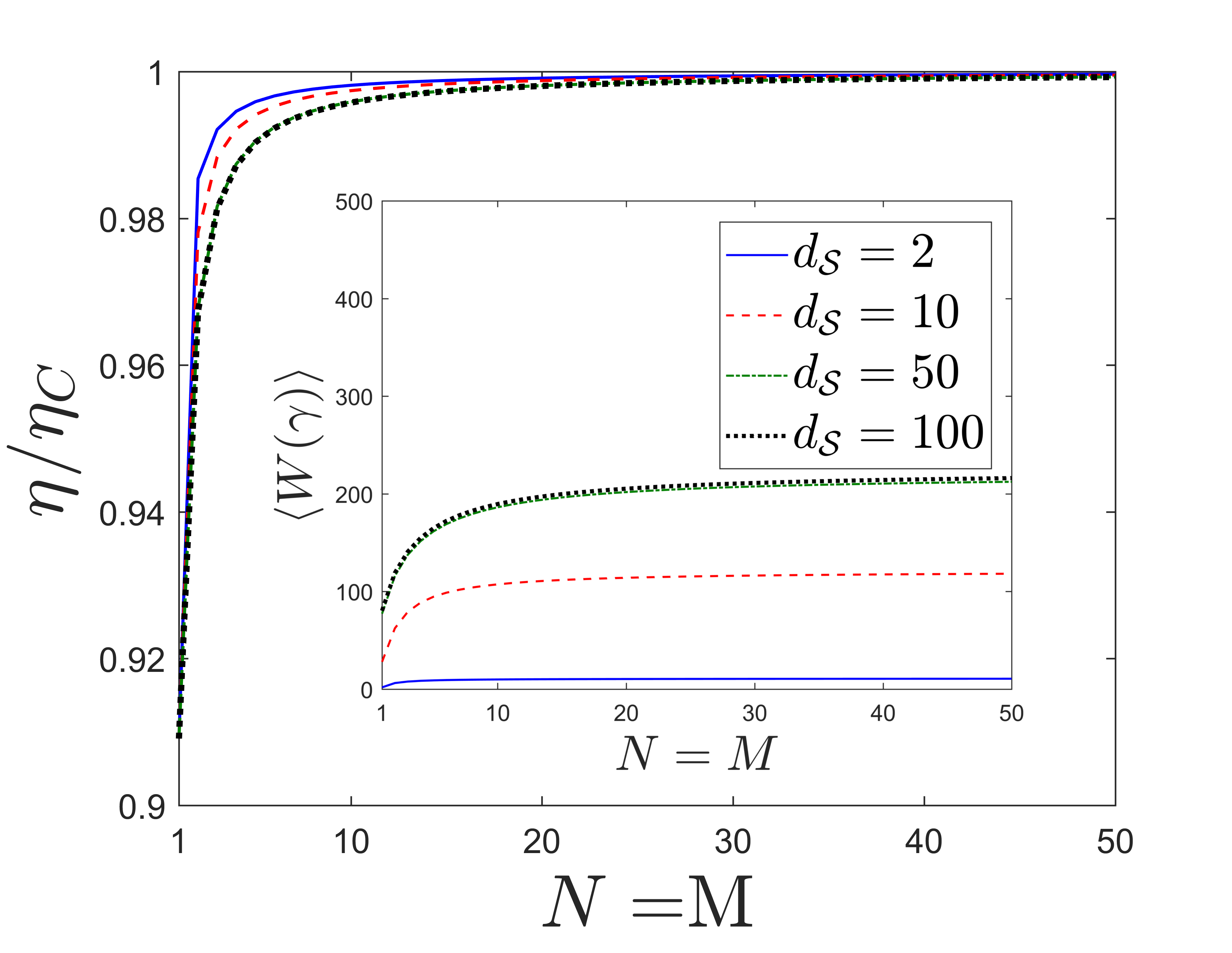}
\caption{Efficiency and average work of a number-conserving cyclic heat engine.  The plots are obtained for $N=M$,  $\beta\sub{\rr} = 1$, $\beta\sub{\e}=10^{-2}$, $\omega\sub{N} = 10$, and $\Omega\sub{M} = 1$. Both the  average work $\<W(\gamma) \>$ and the efficiency $\eta$ increase with $N=M$. For a given $N=M$, the average work (efficiency) increases (decreases) with the  dimension of the working substance,  $d\sub{\s}$. The behaviour seen here will be qualitatively the same even when $N \ne M$; both the average work yield, and the efficiency, will increase as both $N$ and $M$ are made to grow larger, even if they do not do so in unison.  }\label{fig:average-work-efficency}
\end{figure}

Equation \eqref{eq:avg-efficiency-oscillator} states that the only way in which a cyclic heat engine can operate close to the Carnot efficiency is if the relative entropy terms $D\sub{\e}$ and $D\sub{\rr}$  become vanishingly small. In the previous section we saw that the only way this is possible with a small number of bath ensembles is if the engine operates trivially, with $\Delta S = 0$. However, as shown in \eq{eq:avg-work-oscillator} such an engine cannot produce positive work. We shall now see that it is possible to take the terms $D\sub{\e}$ and $D\sub{\rr}$ arbitrarily close to zero, while still obtaining a positive $\Delta S$ and, hence, work extraction, if we use a large number of bath ensembles. This is made possible because a large number of bath ensembles allows the bath frequencies  $\omega\sub{n}$ and $\Omega\sub{m}$ to change smoothly, thus allowing for the engine to operate in the quasistatic limit \cite{Anders-discrete-thermo}.  

Let us therefore define the bath particle frequencies as
\begin{align}\label{eq:sequential-freq}
\omega\sub{n} &:= \omega\sub{0} + \frac{n(\omega\sub{N} - \omega\sub{0})}{N}, \nonumber \\
\Omega\sub{m} &:= \Omega\sub{0} + \frac{m(\Omega\sub{M} - \Omega\sub{0})}{M},
\end{align}
such that $\omega\sub{0} := (\beta\sub{\rr}/\beta\sub{\e})\Omega\sub{M}$ and $\Omega\sub{0} := (\beta\sub{\e}/\beta\sub{\rr})\omega\sub{N}$. Given the previously established constraint of $\beta\sub{\e} \omega\sub{N} < \beta\sub{\rr} \Omega\sub{M}$, \eq{eq:sequential-freq} results in $\omega\sub{n}$ to linearly decrease as $n$ runs from $1$ to $N$, while $\Omega\sub{m}$ linearly increases as $m$ runs from $1$ to $M$. It is clear that for large $N$, $\omega\sub{n} - \omega\sub{n-1}$ becomes vanishingly small, which ensures that the terms $D[\rho\sub{\s}^{\e\sub{n-1}}\|\rho\sub{\s}^{\e\sub{n}}]$ in $D\sub{\e}$ also vanish. The same holds true for the corresponding terms in $D\sub{\rr}$ as $M$ grows large. What is more crucial is that, when both $N$ and $M$ are large, we have $\omega\sub{1} \approx \omega\sub{0}$ and $\Omega\sub{1} \approx \Omega\sub{0}$. This ensures that the remaining terms in $D\sub{\e}$ and $D\sub{\rr}$, namely,  $ D[\rho\sub{\s}^{\rr\sub{M}}\|\rho\sub{\s}^{\e\sub{1}}]$ and $D[\rho\sub{\s}^{\e\sub{N}}\|\rho\sub{\s}^{\rr\sub{1}}] $, will also become vanishingly small.  Note that $\Delta S = S(\rho\sub{\s}^{\e\sub{N}}) - S(\rho\sub{\s}^{\rr\sub{M}})$ is independent of $N$ and $M$. Consequently, given a sufficiently large $N$ and $M$, it is possible for the heat engine to produce a positive amount of work per cycle, with an efficiency that is arbitrarily close to the Carnot value.

In \fig{fig:average-work-efficency} we numerically evaluate the average work and efficiency given by the frequency profiles of \eq{eq:sequential-freq}, and see how the average work and efficiency are affected by the magnitude of $N$ and $M$. Although these parameters are independent of one another, resulting in an improvement in both the work yield and efficiency as they grow larger, for the sake of simplicity we shall assume that  $N=M$.  We see that, as expected, both the average work and efficiency increase with $N=M$, with the efficiency eventually reaching the Carnot value. Moreover, given a fixed value of $N=M$, the average work also increases with the dimension of the working substance, $d\sub{\s}$, converging to a fixed value as $d\sub{\s} \to \infty$. This is because given a Gibbs state $\rho\sub{\s}^{\e\sub{n}}$, as defined in \eq{eq:pseudo-thermal-state}, $S(\rho\sub{\s}^{\e\sub{n}})$ increases with $d\sub{\s}$, converging as
\begin{align}
\lim_{d\sub{\s} \to \infty} S(\rho\sub{\s}^{\e\sub{n}}) = \frac{\beta\sub{\e} \omega\sub{n}}{e^{\beta\sub{\e} \omega\sub{n}} - 1} +\ln{  \frac{e^{\beta\sub{\e} \omega\sub{n}}}{e^{\beta\sub{\e} \omega\sub{n}} - 1}  }.
\end{align}
However, for a fixed value of $N=M$, increasing $d\sub{\s}$ also causes $D\sub{\e}$ and $D\sub{\rr}$ to grow larger, resulting in the  efficiency to decrease. This is consistent with the known power-efficiency trade-offs.

\section{Dimension of thermal baths and the engine's performance}\label{Dimension of thermal baths and the engine's performance}

In \sect{section:general}, we addressed how the dimension of the thermal baths used in a cyclic heat engine limit its efficiency, as per \eq{eq:Efficiency-finite-size-General}. To return to such an analysis with our specific model, we must consider  the Hilbert space dimension of the total baths, as opposed to the dimension of the individual particles that comprise them. Recall that, during each cycle, the working substance $\s$ interacts with $\alpha\sub{n}$ ($\alpha\sub{m}$) particles from the hot (cold) bath ensemble $\e\sub{n}$ ($\rr\sub{m}$). Given that each of these particles has the dimension $d\sub{\e}$ ($d\sub{\rr}$), it follows that the effective dimensions of the total baths  involved are
\begin{align}
\dim(\h\sub{\e}) &= d\sub{\e}^{\sum_{n=1}^N \alpha\sub{n}}, \nonumber \\
\dim(\h\sub{\rr}) &= d\sub{\rr}^{\sum_{m=1}^M \alpha\sub{m}}.
\end{align}
Note that $\dim(\h\sub{\e})$ and $\dim(\h\sub{\rr})$ are to be understood as the size of the effective baths within a single cycle of the engine; the actual baths may indeed be infinitely large, but since only a small part of these are involved during a single cycle of the engine's operation, it is only these dimensions that are pertinent to our considerations. There are two ways in which the effective bath dimensions can be large: either the individual particles have a large dimension, or there are a large number of such particles involved during the engine's cycle. However, as shown above, it is only the second of these that affects the efficiency and work output of the engine. Indeed, the size of the bath particle dimensions $d\sub{\e}$ and $d\sub{\rr}$ do not directly appear in any of the analysis we performed above. However, this does not mean that the particle dimensions do not affect the performance of the engine at all. 

Let us consider the optimal scenario requiring the minimal number of interactions. This is achieved when $d\sub{\e} = d\sub{\rr} = d\sub{\s}$, and the number conserving unitary interaction  for each collision is a SWAP operator.  As such,  we only require a single interaction per ensemble, and so the bath dimensions reduce to $\dim(\h\sub{\e}) = d\sub{\s}^N$ and $\dim(\h\sub{\rr}) =d\sub{\s}^M$. Since the efficiency of the engine grows with $N$ and $M$, while the work ourput increases with $d\sub{\s}$, it follows that increasing the efficiency or power of the engine results in an increase in the dimension of the thermal baths.   In the following section, we shall see how the interplay between bath particle dimension, system dimension, and the effective bath  dimensions becomes much  richer when our interactions are no longer SWAP operations.  

\subsection{Pseudo-thermalization with a Jaynes-Cummings interaction}

In the preceding section we discussed how it is possible to pseudo-thermalize the system to the desired states required by the engine with only one interaction per bath ensemble, resulting in the smallest possible effective bath dimensions. This procedure relied on the ability to perform a SWAP operation between $\s$ and each bath particle. However, it is not always practically possible to perform such a feat. In many situations, we have a very limited way of controlling the interaction between two quantum systems. The paradigmatic example of a number-conserving interaction between $\s$ and the bath particles $X \in \{\e, \rr\}$ is the Jaynes-Cummings interaction Hamiltonian, given as
\begin{align}\label{eq:flip-flop}
V(t):= J(t)(\sigma\sub{\s}^- \otimes \sigma\sub{X}^+ + \sigma\sub{\s}^+ \otimes \sigma\sub{X}^-),
\end{align}
where for  $Y \in \{\s, \e,\rr\}$,
\begin{align}
\sigma\sub{Y}^+ = \sum_{k=0}^{d\sub{Y} -2} \sqrt{k+1}|k+1\>\<k| = (\sigma\sub{Y}^-)^\dagger,
\end{align} 
 and $J(t)$ is the interaction strength which, for  $t \in [0, T_\mathrm{int}]$, equals $J>0$,  and vanishes at all other times.  The Jaynes-Cummings interaction Hamiltonian generally describes the interaction between two systems obeying the Rotating Wave Approximation (RWA), valid when the two systems are in resonance. Examples of systems with such an interaction are an atom coupled to a harmonic oscillator, or two spins interacting via flip-flop processes.  

Since we have modeled our bath particles as a (truncated) harmonic oscillator,  to ensure that the RWA can be made, we shall bring the system Hamiltonian $H\sub{\s}$ in resonance with that of the bath particles before they interact via \eq{eq:flip-flop}. Namely, we shall  set the system Hamiltonian as $H\sub{\s} = \omega\sub{n}\nn\sub{\s}$ when it interacts with particles from the ensemble $\e\sub{n}$, and so forth. We note that since both the state of the system, and its Hamiltonian, will be the same at the start and end of the cycle (specifically, $\rho\sub{\s}^{\rr\sub{M}}$ and  $H\sub{\s} = \Omega_M \nn\sub{\s}$), such quenching of the Hamiltonian will result in a net zero change in internal energy, and it can be ignored; as before, we shall  consider only the heat exchanged with the bath particles, resulting in the average work and efficiency to be given by \eq{eq:avg-work-oscillator} and \eq{eq:avg-efficiency-oscillator} respectively.  Finally, as the reduced state of $\s$ and each bath particle always commutes with its local Hamiltonian,  for simplicity we may  consider only the unitary $U= e^{-i T_\mathrm{int} V }$. While this unitary operator will reduce to a swap operator when $d\sub{\s} = d\sub{\e}=d\sub{\rr} = 2$, and $T_\mathrm{int} = \pi/2J$, this will no longer be possible when $d\sub{\s} >2$.  As such, if we wish for a larger work yield by using larger  dimensions for the working substance, as illustrated by \fig{fig:average-work-efficency}, we will need many interactions per bath ensemble. Consequently, it is unclear how the effective bath dimensions required to achieve a given efficiency, and the average power of such an engine (work yield divided by time, or, number of interactions) will depend on the dimension of the system.  

To answer these questions, we  numerically simulate the engine cycle, fixing the interaction time with each particle as $T_\mathrm{int} = \pi/2J$, and for simplicity restricting the dimension of the hot and cold bath particles to be the same, i.e.,   $d\sub{\e} = d\sub{\rr}$. Moreover, we shall set the other parameters as $\beta\sub{\e} = 10^{-2}$, $\beta\sub{\rr}=1$, $\omega\sub{N}=10$, $\Omega\sub{M}=1$, and $N=M = 10$ throughout. Moreover,  to ensure that the engine  provides the same values of average work and efficiency as given by \eq{eq:avg-work-oscillator} and \eq{eq:avg-efficiency-oscillator}, we shall use the smallest number of interactions $\alpha_n$ ($\alpha_m$) so as to ensure that, during the first run of the engine's cycle, the trace distance between the state of $\s$ after it interacts with ensemble $\e\sub{n}$ ($\rr\sub{m}$), and the pseudo-thermal state $\rho\sub{\s}^{\e\sub{n}}$ ($\rho\sub{\s}^{\rr\sub{m}}$)  defined in \eq{eq:pseudo-thermal-state}, is smaller than $\epsilon = 10^{-9}$. This means that the trace distance between the initial and final state of the engine during its first cycle will be smaller than $\epsilon$.

 \begin{figure}[!htb]
\includegraphics[width = 0.45\textwidth]{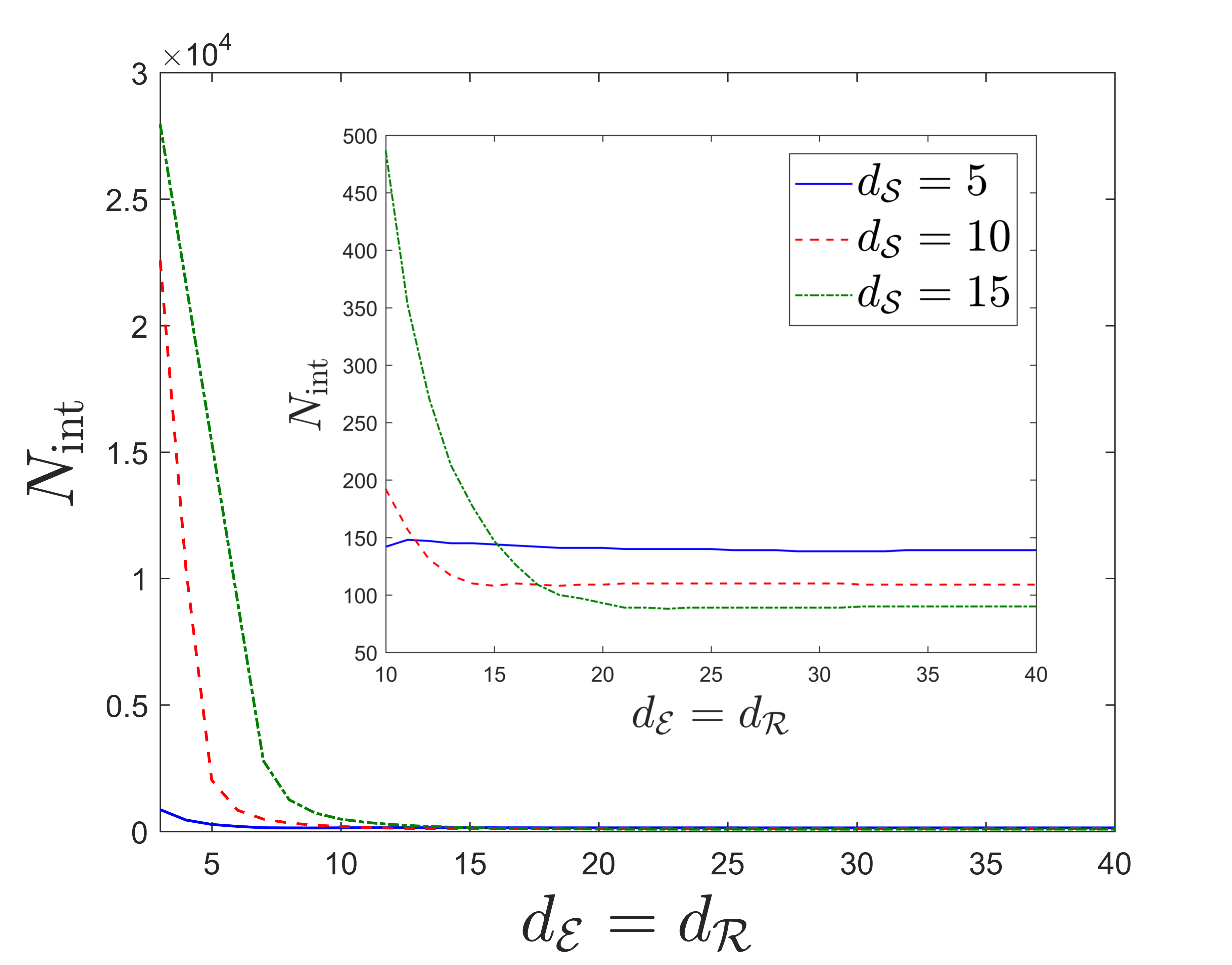}
\caption{ The required  total number of interactions between the working substance $\s$ and particles from the ensembles $\e\sub{n}$ and $\rr\sub{m}$, $N_\mathrm{int} = \sum_n \alpha_n + \sum_m \alpha_m$,  to achieve a pseudo-thermalization parameter of $\epsilon = 10^{-9}$ during the first cycle of the engine, as a function of the bath particle dimension $d\sub{\e}=d\sub{\rr}$. Here we set   $\beta\sub{\e} = 10^{-2}$, $\beta\sub{\rr}=1$, $\omega\sub{N}=10$, $\Omega\sub{M}=1$, and $N=M = 10$.    For a given dimension of the working substance $d\sub{\s}>2$, the total interaction number decreases as the bath particle dimension $d\sub{\e}= d\sub{\rr}$ increases.   }\label{fig:N=M=10-dA-Nint}
\end{figure}

For now, let us consider only the first cycle of the engine. We shall return to the question of many repetitions of the cycle later. In \fig{fig:N=M=10-dA-Nint} we determine the  total number of interactions $N_\mathrm{int} =  \sum_n \alpha\sub{n} + \sum_m \alpha\sub{m}$ required to achieve $\epsilon$-cyclicity given a fixed value of $d\sub{\s}$, and see how this changes with the bath particle dimensions $d\sub{\e}=d\sub{\rr}$. When $d\sub{\e}=d\sub{\rr} < d\sub{\s}$, a large number of interactions  is required, with this number increasing with $d\sub{\s}$. However, the required number of interactions (generally) decreases with $d\sub{\e}$, plateauing as $d\sub{\e}$ becomes very large. Interestingly, as shown by the inset of this figure,  the stabilised interaction number for large $d\sub{\e}$ appears to decrease as $d\sub{\s}$ increases. This suggests that, provided a sufficiently large $d\sub{\e}=d\sub{\rr}$, by increasing $d\sub{\s}$ the power of the engine (average work per interaction) will also increase.

\begin{figure}[!htb]
\includegraphics[width = 0.45\textwidth]{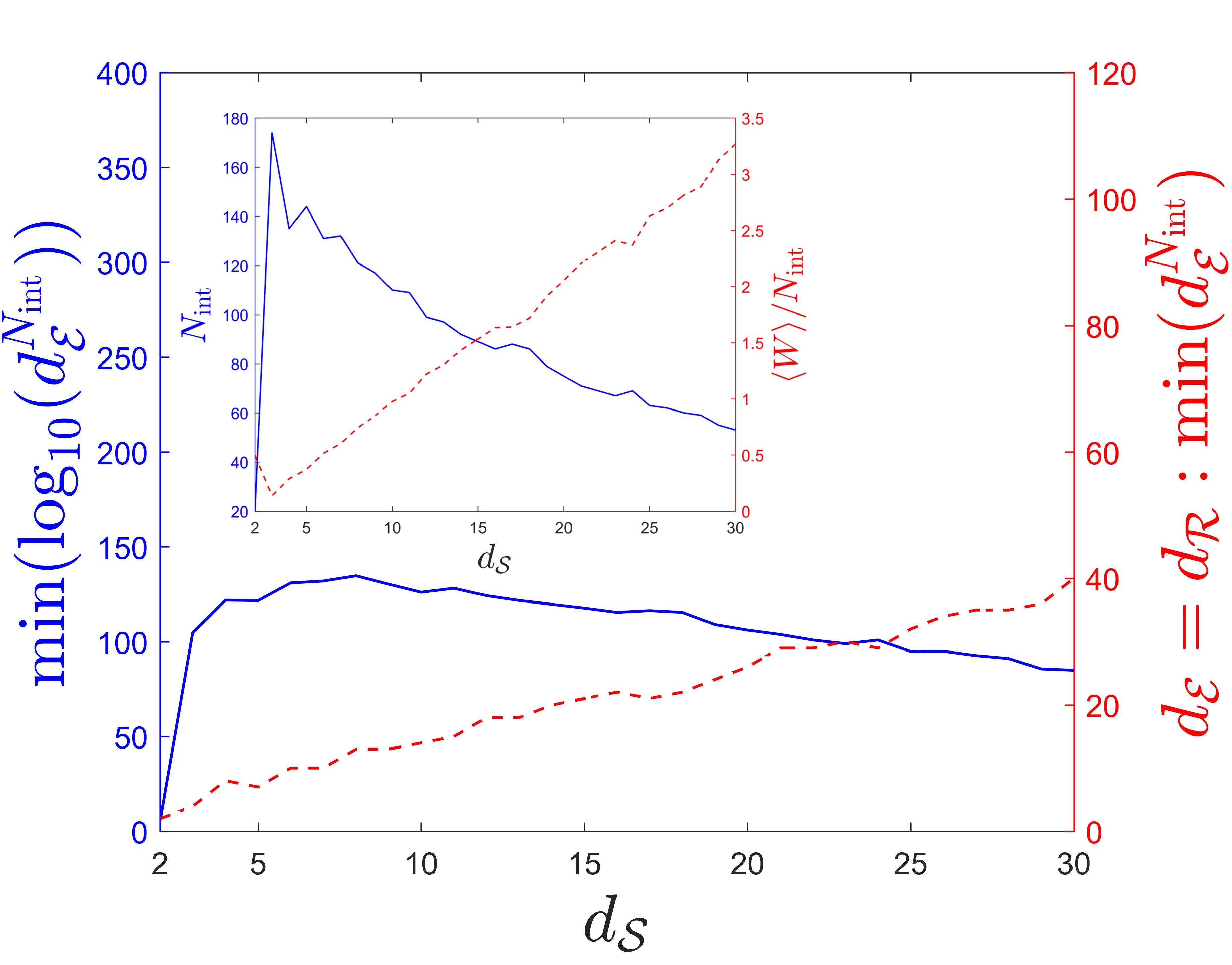}
\caption{ The optimal choice for bath particle dimensions $d\sub{\e} = d\sub{\rr}$ so as to minimise the dimension of the full effective baths $\dim(\h\sub{\e}\otimes \h\sub{\rr}) = d\sub{\e}^{N_\mathrm{int}}$, as a function of the dimension of the working substance $d\sub{\s}$. The pseudo-thermalisation parameter is set to $\epsilon = 10^{-9}$, while   $\beta\sub{\e} = 10^{-2}$, $\beta\sub{\rr}=1$, $\omega\sub{N}=10$, $\Omega\sub{M}=1$, and $N=M = 10$.      The efficiency of this engine is $\eta/\eta\sub{C} \approx 0.99$ which, as shown in \fig{fig:average-work-efficency}, decreases as $d\sub{\s}$ increases. As $d\sub{\s}>2$ increases, the dimension of the effective baths is minimised by increasing the particle dimensions $d\sub{\e}$, allowing for the total number of interactions $N_\mathrm{int}$ to decrease. This allows for a greater power output. Moreover, for a sufficiently large $d\sub{\s}$, the optimal dimension of the effective baths decreases with $d\sub{\s}$.  }\label{fig:N=M=10-d_S-Min_bath_dim}
\end{figure}

Of course, although using larger  particle dimensions may result in a decrease in the number of interactions $N_\mathrm{int}$, this may come at the expense of a larger dimension of the effective total bath, which is $\dim(\h\sub{\e} \otimes \h\sub{\rr}) = d\sub{\e}^{N_\mathrm{int}}$.  Therefore, for a given dimension $d\sub{\s}$, we wish to determine the optimal choice of $d\sub{\e} = d\sub{\rr}$ so that this effective bath dimension will be minimised. This is shown in \fig{fig:N=M=10-d_S-Min_bath_dim}. In the main figure, the blue solid line depicts the smallest effective dimension of the total bath, $d\sub{\e}^{N_\mathrm{int}}$, as a function of system dimension $d\sub{\s}$. The red dotted line, meanwhile, shows the particle dimensions $d\sub{\e}=d\sub{\rr}$ that achieve this optimal bath dimension. The inset of the figure shows the total number of interactions $N_\mathrm{int}$, and the resulting power $\<W\>/N_\mathrm{int}$ (average work per interaction), achieved by such particle dimensions chosen to minimise the size of the full baths. As we can see, the smallest overall bath dimension is achieved when $d\sub{\s} = 2$, where given a choice of $d\sub{\e} = d\sub{\rr} = 2$,  $U$ reduces to a SWAP operation and only one particle per ensemble is required. As $d\sub{\s}$ grows larger, however, $U$ does not reduce to a SWAP map, and in general many interactions will be needed to achieve pseudo-thermalization. Recall from \fig{fig:N=M=10-dA-Nint} that as $d\sub{\s}$ increases, a larger particle dimension $d\sub{\e}$ is required to minimise the number of interactions $N_\mathrm{int}$, while $N_\mathrm{int}$ (given a sufficiently large $d\sub{\e}$) decreases with $d\sub{\s}$. Therefore, as shown in \fig{fig:N=M=10-d_S-Min_bath_dim}, the optimal bath particle dimensions $d\sub{\e}$ (generally) increase with $d\sub{\s}$, while for $d\sub{\s} >2$, the optimal number of interactions $N_\mathrm{int}$ (generally) decrease with $d\sub{\s}$. The rate at which $d\sub{\e}=d\sub{\rr}$  ($N_\mathrm{int}$) increases (decreases) with $d\sub{\s}$ results in $d\sub{\e}^{N_\mathrm{int}}$ to grow with $d\sub{\s}$ when $d\sub{\s}$ is small, while it decreases with $d\sub{\s}$ when $d\sub{\s}$ is large.

 \begin{figure}[!htb]
\subfigure[]{\includegraphics[width = 0.45\textwidth]{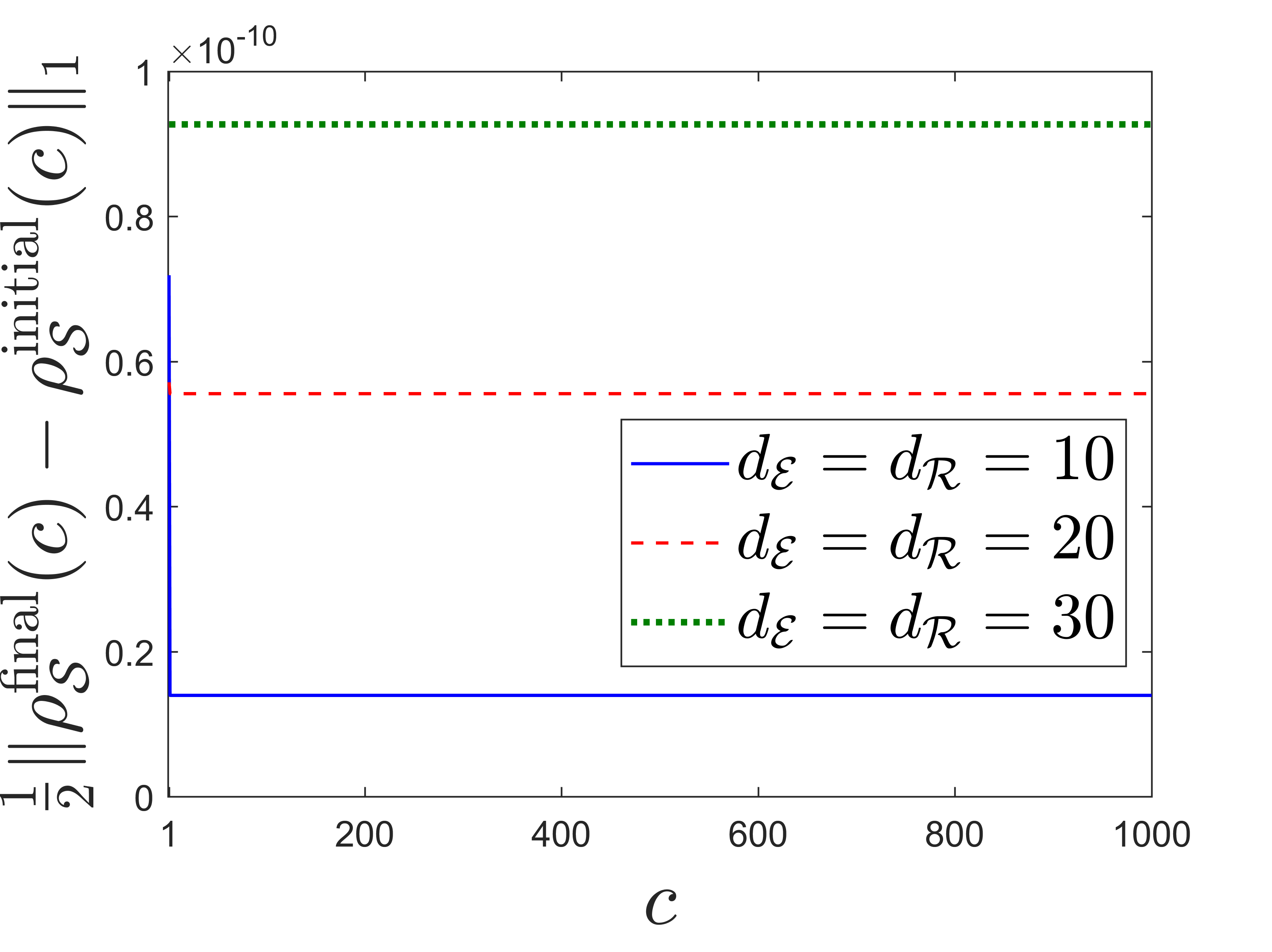} \label{fig:Many-cycles-cyclicity}}
\subfigure[]{\includegraphics[width = 0.45\textwidth]{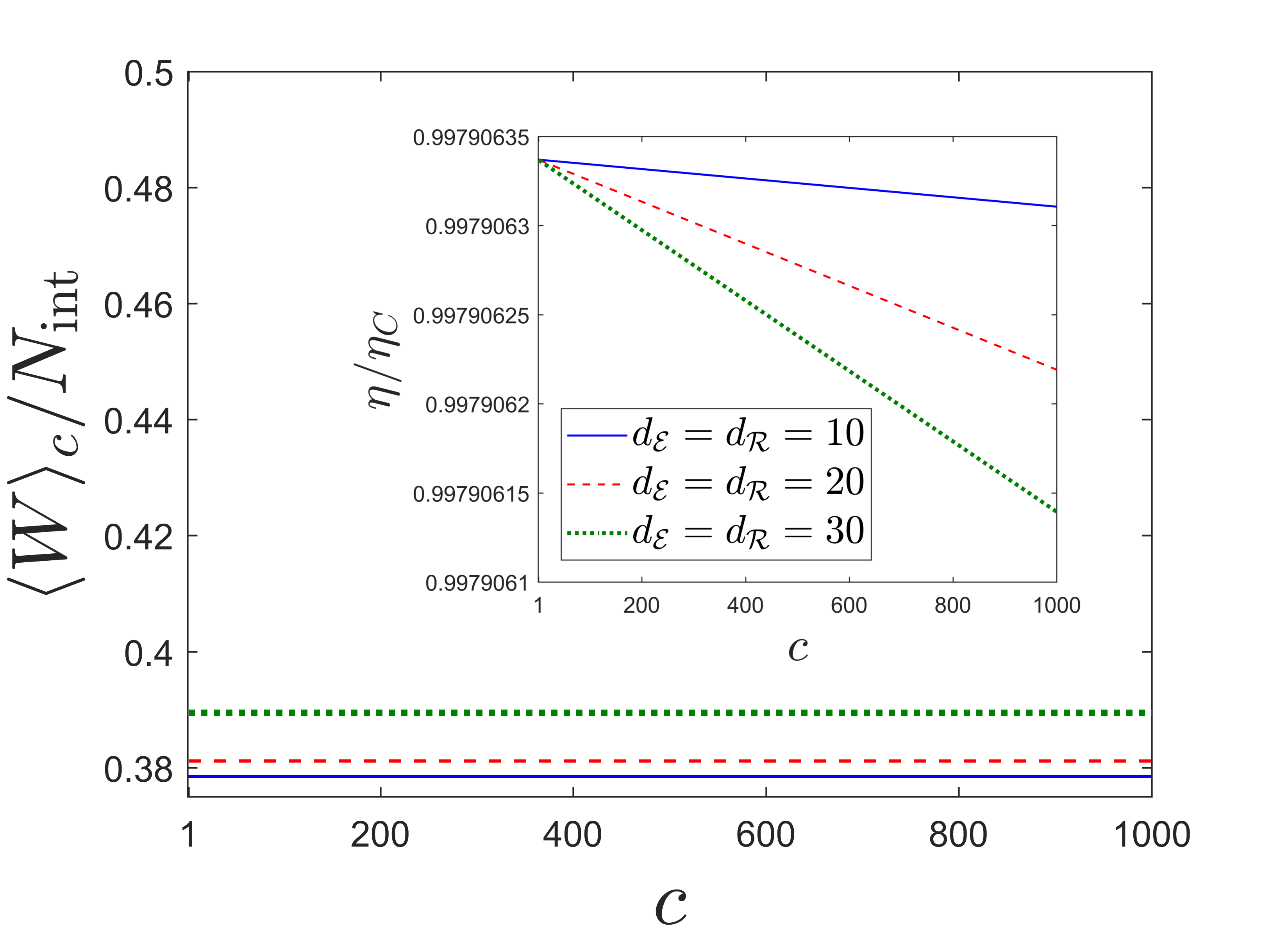} \label{fig:Many-cycles-power-efficiency}}
\caption{Performance of the engine after $c$ repetitions of the cycle. Here we set $d\sub{\s}=5$, $\beta\sub{\e} = 10^{-2}$, $\beta\sub{\rr}=1$, $\omega\sub{N}=10$, $\Omega\sub{M}=1$, and $N=M = 10$. The initial and final states of $\s$ during cycle $c$ are denoted as $\rho\sub{\s}^\mathrm{initial}(c)$ and $\rho\sub{\s}^\mathrm{final}(c)$, respectively, where $\rho\sub{\s}^\mathrm{initial}(1) = \rho\sub{\s}^{\rr\sub{M}}$ and $\rho\sub{\s}^\mathrm{final}(c) = \rho\sub{\s}^\mathrm{initial}(c+1)$. The number of interactions $\alpha_n$ and $\alpha_m$ are fixed for each cycle, and chosen to achieve $\epsilon$-pseudo-thermalisation with the bath ensembles during the first cycle. (a) The cyclicity of the engine, characterised as the trace distance between the initial and final state of $\s$ for each cycle $c$. This remains within $\epsilon$, but the engine becomes less cyclic as the bath particle's dimension increases. (b) The integrated average work is defined as $\<W\>_c := \sum_{i=1}^c \<W^i\>/c$, where $\<W^i\>$ is the average work evaluated for cycle $i$. While $\<W\>_c$ decreases with $c$, this is negligible, and the average integrated power $\<W\>_c/N_\mathrm{int}$ still increases with bath particle dimension. In the inset, we see that the efficiency  decreases with repetitions of the cycle, with the rate being faster when the bath particle dimensions increase.    }\label{fig:Many-cycles}
\end{figure}

Now let us return to the question of repeated cycles. If we fix the number of interactions $\alpha_n$ and $\alpha_m$ for every cycle, then  the final state of $\s$ after $c$ cycles, $\rho\sub{\s}^\mathrm{final}(c)$, will diverge from the initial state $\rho\sub{\s}^{\rr\sub{M}}$. However, this does not mean that the engine will cease to be $\epsilon$-cyclic, since $\rho\sub{\s}^\mathrm{final}(c)$ will still remain within $\epsilon$ to $\rho\sub{\s}^\mathrm{initial}(c)$, namely, the initial state of $\s$ during the $c$\ts{th} cycle.  This is shown in \fig{fig:Many-cycles-cyclicity}. Interestingly, however, the engine becomes less cyclic as the bath particle dimension increases. Moreover, for a finite $\epsilon$ the engine will degrade after many cycles. Specifically, both the total average power, and the efficiency, decrease with repetitions of the cycle. However, for the case of $\epsilon = 10^{-9}$ this degradation is negligible as shown in \fig{fig:Many-cycles-power-efficiency}. Here, $\<W\>_c$ is defined as the average integrated work after $c$ cycles. Interestingly, we see that while increasing the bath particle dimension results in the average power to increase even after a thousand cycles (due to the fewer number of interactions per cycle), a larger bath particle dimension results in the efficiency of the engine to degrade at a faster rate.

In conclusion, when $U$ can always be constructed as a SWAP operator, given a fixed value of $N$ and $M$, increasing the dimension of the working substance results in both the power of the engine, and the dimension of the baths, to increase, while the efficiency decreases.   However, when $U$ is generated by a Jaynes-Cummings interaction, when $d\sub{\s}$ is sufficiently large, its increase will result in the power to increase while both the bath dimension and efficiency decrease.  Moreover,  when $U$ is a swap operator  the engine is completely cyclic, with $\epsilon = 0$. Consequently, the engine's performance will not degrade with repeated cycles. However, with the Jaynes-cummings interaction for $d\sub{\s} >2$, the engine cannot be fully cyclic, and so for a finite number of interactions per cycle, the engine's performance will degrade with time.

\section{Achieving the stochastic Carnot efficiency with finite-dimensional baths}\label{Achieving the stochastic Carnot efficiency with finite-dimensional baths}

Thus far, we have seen that a cyclic heat engine can  achieve the Carnot efficiency, determined as the ratio of average work with respect to the average heat absorbed from the hot bath, only when both the hot and cold baths have an infinite-dimensional Hilbert space. This is true even when the individual bath particles have a small size, since the Carnot efficiency will require an infinite number of them as dictated by the number of ensembles $N$ and $M$. However, we shall now show that the maximum likelihood stochastic efficiency can approach the Carnot efficiency when the baths are truly finite dimensional.  Specifically, the efficiency of the most likely trajectory will approach the Carnot value if we  take only the dimension of the cold bath to infinity, while the stochastic efficiency that occurs with the largest probability approaches the Carnot value even when both the hot and cold baths have a finite dimension. We note that such efficiencies are  evaluated only for a single cycle of the engine, and are not to be confused with the stochastic efficiencies in the long time limit, evaluated after many repetitions of the cycle. 

\begin{figure}[!htb]
\includegraphics[width =0.45\textwidth]{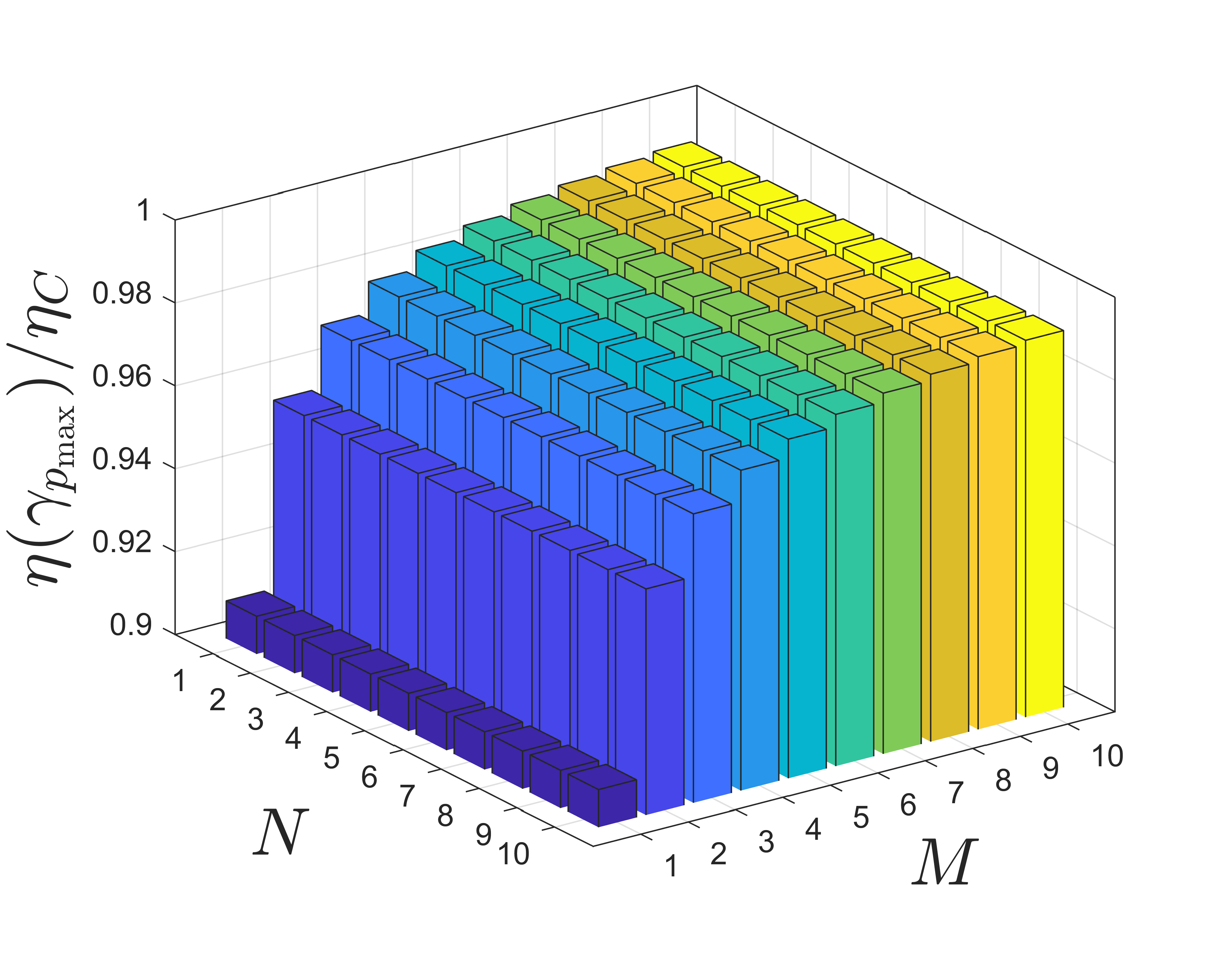}
\caption{The efficiency of the most likely trajectory, $\eta(\gamma_{p_\mathrm{max}})$ as a function of the number of hot and cold bath ensembles $N$ and $M$. The parameters are set as $d\sub{\s} = 2$,  $\beta\sub{\rr} = 1$, $\beta\sub{\e}=10^{-2}$,  $\omega\sub{N} = 10$ and $\Omega\sub{M} =  1$.  $\eta(\gamma_{p_\mathrm{max}})$ monotonically increases with $M$, approaching the Carnot efficiency. The result does not depend on the number of hot bath ensembles $N$.} \label{fig:Efficiency_Max_Prob_Traj}
\end{figure}

Using \eq{eq:heat-traj} and \eq{eq:sequential-freq}, the heat contributions for each trajectory of the above protocol can be written as $Q\sub{\e}(\gamma) = \Delta \e (\gamma)/N r\sub{\beta}$ and $Q\sub{\rr}(\gamma) = \Delta \rr(\gamma)/M$, where $r\sub{\beta} := \beta\sub{\e}/\beta\sub{\rr}$ and
\begin{align}\label{eq:traj-heat-discretefreq}
\Delta \e (\gamma)& := \Omega\sub{M}(\sigma\sub{\e}^\gamma  - N j_0) - \omega\sub{N} r\sub{\beta}(\sigma\sub{\e}^\gamma  - N j_N), \nonumber \\
\Delta \rr (\gamma)& := \Omega\sub{M}(\sigma\sub{\rr}^\gamma  - M k_M) - \omega\sub{N} r\sub{\beta}(\sigma\sub{\rr}^\gamma  - M j_N),
\end{align}
where  $\sigma\sub{\e}^\gamma := \sum_{n=0}^{N-1}j_n$,   $\sigma\sub{\rr}^\gamma := \sum_{m=0}^{M-1}k_m$ and, as assumed before, $k_0 = j_N$. The efficiency for each trajectory (such that both $W(\gamma)>0$ and $Q\sub{\e}(\gamma)>0$) will thus be given by \eq{eq:efficiency-traj} and \eq{eq:traj-heat-discretefreq} to be
\begin{align}\label{eq:traj-efficiency-discretefreq}
\eta(\gamma) = 1 - r\sub{\beta} \frac{N }{M}\left(\frac{\Delta \rr (\gamma) - M \Delta E(\gamma)}{\Delta \e (\gamma)}\right),
\end{align}
where we recall that $\Delta E(\gamma):= \<j_0|H\sub{\s} |j_0\> - \<k_M|H\sub{\s} |k_M\>$ is the change in internal energy of $\s$ during each trajectory. Since the probability for each trajectory, in the limit of $\epsilon \to 0$, is given by \eq{eq:prob-traj}, i.e., a product of the probabilities that the pseudo-thermal state $\rho\sub{\s}^{\e\sub{n}}$ occupies the state $\ket{j_n}$ and so on,  it follows that the trajectory that occurs with the highest probability is one where $k_m=0$ for all $m$, and $j_n=0$ for all $n$. However, the resulting work and heat values for this trajectory will both be zero, and so this trajectory does not have a well-defined efficiency. Consequently, we must have $j_n=1$ for at least one value of $n$. Since $\omega\sub{N}$ is the smallest frequency from the set $\{\omega\sub{n}\}$, it follows that the most probable trajectory for which an efficiency is defined is one where $j_n=1$ for $n=N$, with all other outcomes being zero.  In other words, the trajectory where $\s$ starts off in the state $\ket{0}$, and absorbs one quanta of energy from the ensemble $\e\sub{N}$, emitting this to the ensemble $\rr\sub{1}$, and remaining in the state $\ket{0}$ thereafter.  The efficiency for this trajectory, which we refer to as $\gamma_{p_\mathrm{max}}$, is obtained from   \eq{eq:traj-efficiency-discretefreq} to be
\begin{align}\label{eq:efficiency-most-likely-traj}
\eta(\gamma_{p_\mathrm{max}}) = \eta\sub{C} - \frac{\Omega\sub{M} - \omega\sub{N} r\sub{\beta}}{M \omega\sub{N}} < \eta\sub{C},
\end{align}
where the inequality follows from the requirement that $\omega\sub{N} \beta\sub{\e} < \Omega\sub{M} \beta\sub{\rr}$. Consequently,   the efficiency of the most probable trajectory of the engine  approaches $\eta\sub{C}$ from below as $M \to \infty$. Interestingly, the number of hot bath ensembles $N$ does not affect this. To see this, we note that $Q\sub{\e}(\gamma_{p_\mathrm{max}}) = \omega\sub{N}$, which is a parameter chosen independently of $N$. On the other hand, $Q\sub{\rr}(\gamma_{p_\mathrm{max}}) = \Omega\sub{1}$ which, by \eq{eq:sequential-freq}, approaches $\Omega\sub{0} = r\sub{\beta} \omega\sub{N}$ in the limit as $M$ goes to infinity. As such, we have $W(\gamma_{p_\mathrm{max}})/Q\sub{\e}(\gamma_{p_\mathrm{max}}) = 1 - \Omega\sub{1}/\omega\sub{N}$, which approaches the Carnot efficiency as  $M \to \infty$. This is shown in \fig{fig:Efficiency_Max_Prob_Traj}, where we report the dependence of the efficiency of the most likely trajectory for general $N$ and $M$.
 
The efficiency of the most likely trajectory is not to be confused with the most likely efficiency. In fact, generally there are multiple trajectories with the same efficiency. As such, we may define the set of unique efficiencies $\tilde \eta$, with probabilities $p(\tilde \eta):= \sum_{\gamma: \eta(\gamma) = \tilde \eta} p(\gamma)$. We normalize these probabilities as $\sum_{\tilde \eta} p(\tilde \eta) =1$, so that only trajectories with a well defined efficiency are accounted for,  i.e. we exclude trajectories for which the efficiency is not defined. The most likely efficiency, which we denote as $\tilde \eta_{p_\mathrm{max}}$, is the one for which $p(\tilde \eta)$ takes the maximal value. 

\begin{figure}[!htb]
\subfigure[]{\includegraphics[width = 0.45\textwidth]{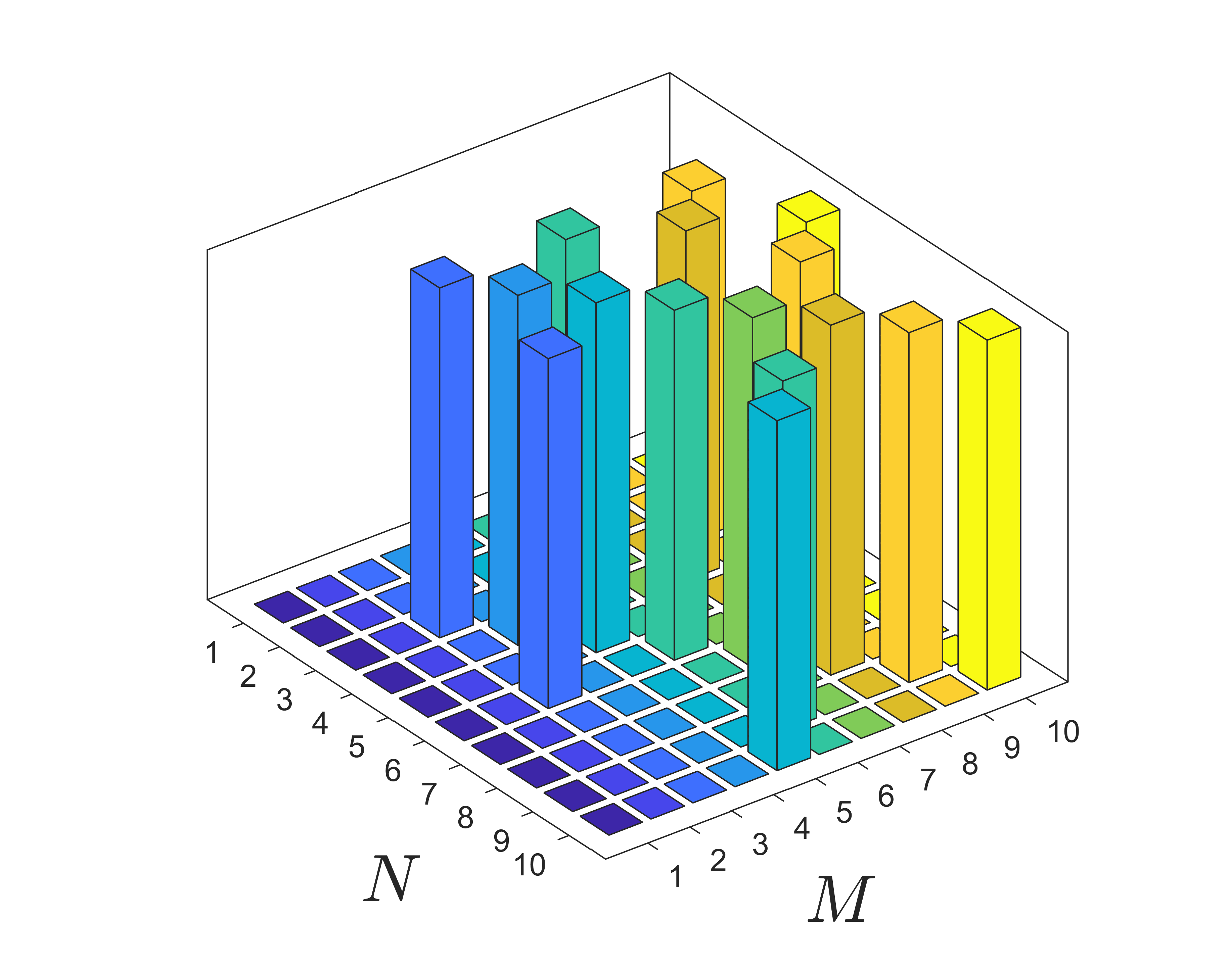} \label{fig:Max-Prob-Eff-Carnot}}
\subfigure[]{\includegraphics[width = 0.45\textwidth]{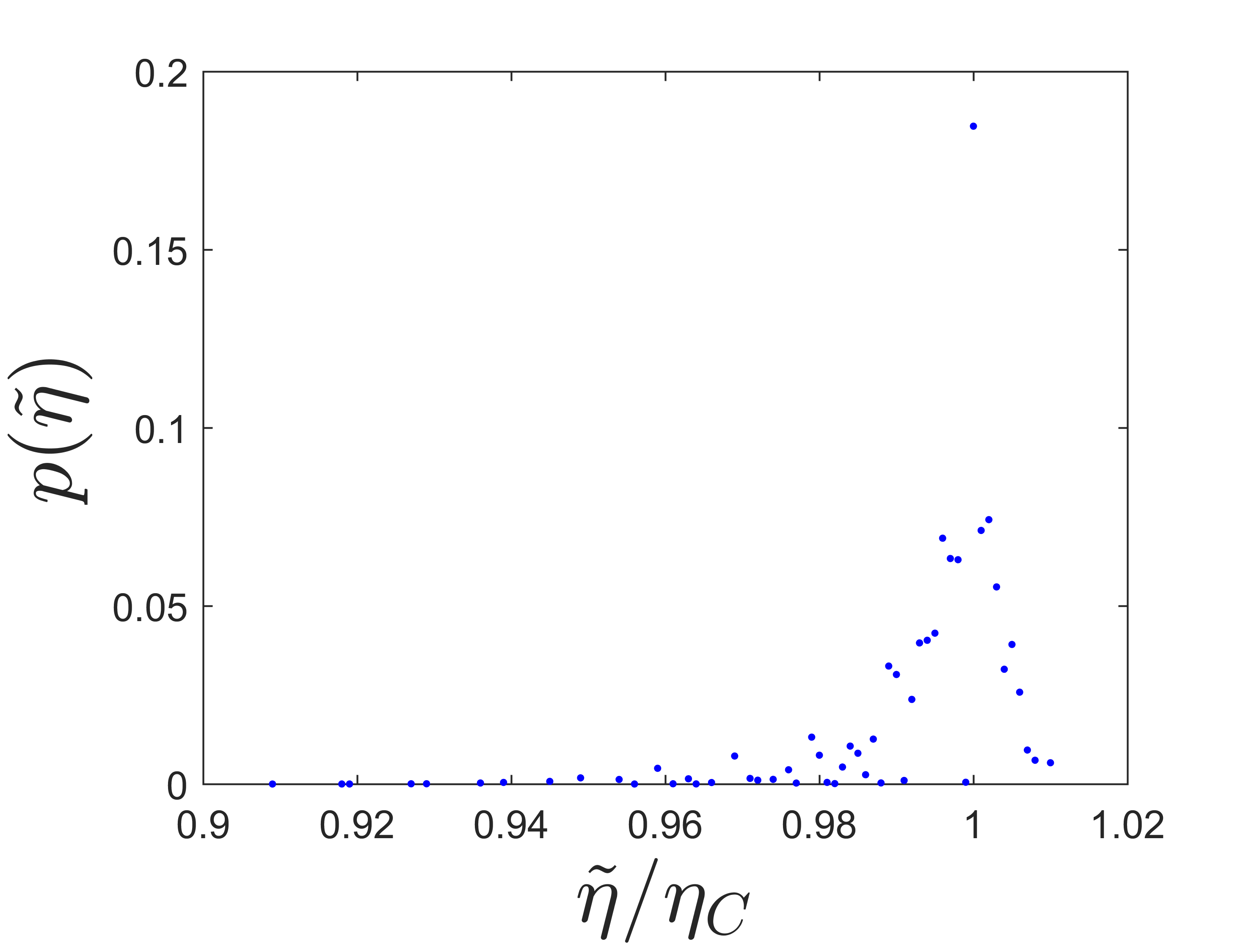} \label{fig:Stochastic_Efficiency}}
\caption{A cyclic heat engine where the working substance is a qubit ($d\sub{\s} = 2$), with the parameters $\beta\sub{\e}=10^{-2}$,  $\beta\sub{\rr} = 1$,  $\omega\sub{N} = 10$ and $\Omega\sub{M} = \<1|H\sub{\s}|1\> - \<0|H\sub{\s}|0\>= 1$.  (a) shows the choice of $N$ and $M$ such that  the most likely efficiency, $\tilde \eta_{p_\mathrm{max}}$, equals the Carnot efficiency $\eta\sub{C}$. (b) shows the full distribution of the stochastic efficiency $\tilde \eta$ for the case of $N=M=10$. This distribution is peaked at the Carnot efficiency.    }\label{fig:Max_Prob_Eff=Carnot}
\end{figure}

From   \eq{eq:traj-efficiency-discretefreq} we see that trajectories satisfying
\begin{align}\label{eq:Carnot-traj-condition}
\frac{N }{M}\left(\frac{\Delta \rr (\gamma) - M \Delta E(\gamma)}{\Delta \e (\gamma)}\right) = 1
\end{align}
will have the Carnot efficiency. The simplest (but not exclusive) case where \eq{eq:Carnot-traj-condition} is satisfied is when $N=M$, with the  trajectories $\gamma$ such that $\sigma\sub{\rr}^\gamma = \sigma\sub{\e}^\gamma$ and  $j_0 = k_M$.  $N=M=2$ are the smallest values that satisfy this; when $N=M=1$, the condition is satisfied only by the trajectory with all outcomes being the same, so the efficiency will not be defined.  However, having $N=M \geqslant 2$ is neither necessary nor sufficient for $\tilde \eta_{p_\mathrm{max}} = \eta\sub{C}$, as this will also depend on the specific choices of the parameters  $\omega\sub{N}$, $\Omega\sub{M}$, $r\sub{\beta}$, and $H\sub{\s}$.  This is shown in \fig{fig:Max-Prob-Eff-Carnot}, which reports the combinations of $N$ and $M$ which result in $\tilde \eta_{p_\mathrm{max}} = \eta\sub{C}$ for $d\sub{\s} = 2$   given a choice of parameters. We see that when $N=M=2$, the most likely efficiency is not the Carnot value, while the most likely efficiency is the Carnot value in some cases where $N \ne M$.  In \fig{fig:Stochastic_Efficiency} we show the full distribution of efficiencies for the case of $N=M=10$. As can be seen, this distribution is peaked at $\tilde\eta = \eta\sub{C}$.

In conclusion, \eq{eq:efficiency-most-likely-traj} demonstrates that the most likely trajectory approaches the Carnot value when only $M$ is taken to infinity, whereas \fig{fig:Max_Prob_Eff=Carnot} shows that the most likely efficiency can be exactly the Carnot value when both $N$ and $M$ are finite. Since the dimension of the thermal baths are determined by $N$ and $M$, and in the simplest case where the collision interactions effect a SWAP operation, are simply given as $\dim(\h\sub{\e}) = d\sub{\s}^N$ and $\dim(\h\sub{\rr}) = d\sub{\s}^M$, it follows that (i) the efficiency of the most likely trajectory, $\eta(\gamma_{p_\mathrm{max}})$, approaches $\eta\sub{C}$ when $\dim(\h\sub{\e}) $ is finite and only $\dim(\h\sub{\rr})$ goes to infinity; and (ii) the most likely efficiency $\tilde \eta_{p_\mathrm{max}}$ can equal $\eta\sub{C}$ when both $\dim(\h\sub{\e}) $ and $\dim(\h\sub{\rr})$ have a finite value. This is in contrast to the efficiency of the average work, which approaches $\eta\sub{C}$ when both $\dim(\h\sub{\e}) $ and $\dim(\h\sub{\rr})$ approach infinity. Of course, we note that our definitions of stochastic efficiency, namely $\eta(\gamma_{p_\mathrm{max}})$ and $\tilde \eta_{p_\mathrm{max}}$, are evaluated for just one cycle of the engine. As was reported in \cite{Verley2014}, the Carnot efficiency is in fact the least likely efficiency when we consider infinitely many repetitions of the engine's cycle, which is irrespective of the dimension of the baths.

\section{Conclusions}

In this work we have analyzed a simple model of a cyclic heat engine operating between a hot and a cold bath of finite size. Each bath consists of a finite set of ensembles of identical particles, each with a finite-dimensional Hilbert space. These particles can be considered as a truncated harmonic oscillator, with the different ensembles being characterised by the oscillator's frequency.  Given a number conserving interaction between the working substance and each of the bath particles, we have shown that the engine can produce positive average work, with an efficiency that approaches the Carnot value as the number of bath ensembles tends to infinity, with the corresponding  frequencies changing smoothly.  This illustrates that, although the individual particles may have a small dimension, the dimension of the total baths must become infinitely large for the efficiency of average work to approach the Carnot Limit. 

Moreover, we showed that when the number conserving interaction between the working substance and the bath particles effects a SWAP operation, then increasing the power (efficiency) of the engine will require a larger particle dimension (particle number), resulting in a larger dimension of the effective baths. In contrast, when the collision is given by a Jaynes-Cummings interaction, it is possible to increase the power output of the engine while decreasing the bath dimensions.  

The proposed engine also allows for a simple characterisation of the stochastic efficiency defined for each cycle of the engine (as opposed to the stochastic efficiency in the long time limit or, equivalently, over infinitely many cycles). In contrast to the consideration of the efficiency of average work, we demonstrated the possibility for the most likely stochastic efficiency to equal the Carnot efficiency even when both hot and cold baths are composed of a finite number of ensembles and, hence, have a finite-size Hilbert space.

\acknowledgments

We are thankful to E. Lutz for inspiring discussions and comments. We acknowledge support from EPSRC 
via Grant EP/P030815/1. A.R. acknowledges the KITP program ``Thermodynamics of quantum systems: Measurement, engines, and control'' 2018, where part of this work has been done. This research was supported in part by the National Science Foundation under Grant No. NSF PHY-1748958.  



\bibliography{Oscillator-Engine-References}

\newpage

\appendix

\makeatletter
\renewcommand\p@subsection{\thesection\,}
\makeatother
\makeatletter
\renewcommand\p@subsubsection{\thesection\,\thesubsection\,}
\makeatother

\section{Properties of the pseudo-thermalization quantum channel}\label{app:stationary state}
Here, we shall discuss some properties of the quantum channel implemented on $\s$ due to a collision with a bath particle $\e$,
\begin{align}
\Lambda_\e : \rho\sub{\s} \mapsto \tr\sub{\bar \s}[U(\rho\sub{\s} \otimes \rho\sub{\e})U^\dagger],
\end{align}
 where $U$ is a number conserving unitary operator on $\h\sub{\s} \otimes \h\sub{\e}$ and  $\rho\sub{\e} = e^{-\beta\sub{\e} \omega\sub{\e} \nn\sub{\e}}/\tr[e^{-\beta\sub{\e} \omega\sub{\e} \nn\sub{\e}}]$ is the thermal state  of the bath particle.  To this end, let us decompose the full Hilbert space as
\begin{align}
 \h\sub{\s}\otimes\h\sub{\e} = \bigoplus_{l=0}^{L} \h_l,
\end{align}
where $L = d\sub{\s} + d\sub{\e} - 2$, and  $\h_l$ is the $l$-number subspace spanned by the vectors $\ket{l-k,k}$ (the left system is $\s$, and the right is $\e$),   and has the dimension
\begin{align}
\mathrm{dim}(\h_l) &= l+1 - \max(0, l - d\sub{\s} + 1), \nonumber \\
&\, \, \,  - \max(0, l - d\sub{\e} + 1) , \nonumber \\
& \equiv 1 + \min(l, d\sub{\s} - 1) - \max(0, l - d\sub{\e} + 1).
\end{align} 
Given that the unitary operator $U$ conserves the total number, we may write it as the direct sum 
\begin{align}
U = \bigoplus_{l=0}^{L} U_l,
\end{align}
where $U_l$ is a unitary operator acting on  $\h_l$.

First, we show that if $\rho\sub{\s}$ commutes with $\nn\sub{\s}$, then so too will $\Lambda_\e(\rho\sub{\s})$. Let the state of $\s$ and $\e$ be $\rho\sub{\s} = \sum_\mu p_\mu \prs{\mu}$ and $\rho\sub{\e} = \sum_\nu q_\nu \pre{\nu}$, respectively, where $\ket{\mu}$ and $\ket{\nu}$ are eigenstates of $\nn\sub{\s}$ and $\nn\sub{\e}$, respectively. The matrix elements of the reduced state of $\s$,  after a number conserving unitary interaction with $\e$, are thus given in the $\nn\sub{\s}$ representation as 
\begin{align}
\<i|\Lambda_{\e}(\rho\sub{\s})|j\> &= \<i| \tr\sub{\aa} [U(\rho\sub{\s} \otimes \rho\sub{\e})U^\dagger]|j\> \nonumber \\
& =\sum_{\mu,\nu,\nu'}p_\mu q_\nu \<i,\nu'| U |\mu,\nu\>\<\mu, \nu|U^\dagger |j,\nu'\>\nonumber \\
&=\sum_{\mu,\nu,\nu'} p_\mu q_\nu \<i,\nu'| U |\mu,\nu\> \<j,\nu'|U|\mu, \nu\>^* .
\end{align}
Given that $U$ conserves the total number, it follows that
\begin{align}
\<i,\nu'| U |\mu,\nu\> \<j,\nu'|U|\mu, \nu\>^* = \delta_{i,j}|\<i,\nu'| U |\mu,\nu\>|^2.
\end{align}  
Consequently, the only non-vanishing matrix elements of $\Lambda_{\e}(\rho\sub{\s})$ are $\<i|\Lambda_{\e}(\rho\sub{\s})|i\>$. In other words, $\Lambda_{\e}(\rho\sub{\s})$ commutes with the number operator.

Now, we show that $\Lambda_\e( \rho\sub{\s}^\e) =  \rho\sub{\s}^\e$, where 
\begin{align}
\rho\sub{\s}^{\e} = \frac{e^{-\beta\sub{\e} \omega\sub{\e} \nn\sub{\s}}}{\tr[e^{-\beta\sub{\e} \omega\sub{\e} \nn\sub{\s}}]}
\end{align}
is the pseudo-thermal state of $\s$ with respect to the observable $\omega\sub{\e}\nn\sub{\s}$, which is not necessarily its Hamiltonian. 
If the compound system $\s + \e$ is prepared in the state $\rho = \rho\sub{\s}^{\e}\otimes \rho\sub{\e}$,  we can write this as $\rho = \sum_l \tilde \rho_l$, where $\tilde \rho_l$ is a sub-normalized state on $\h_l$. It is simple to verify that $\tilde \rho_l \propto \one_l$. This is because $\tilde \rho_l$ is diagonal in the $|l-k, k\>$ basis, and for every $k$,  
\begin{align}
\<l-k,k| \tilde \rho_l |l-k,k\>= \frac{e^{-\beta\sub{\e}\omega\sub{\e} l}}{\tr[e^{-\beta\sub{\e} \omega\sub{\e} \nn\sub{\s}}] \tr[e^{-\beta\sub{\e} \omega\sub{\e} \nn\sub{\e}}]}.
\end{align}
 Consequently, we have
\begin{align}
U \rho U^\dagger = \sum_l U_l \tilde \rho_l U_l^\dagger = \rho.
\end{align}

From this, it is apparent that the pseudo-thermalization of a generic state will be guaranteed if, in addition to number conservation, none of the 
 $U_l$ operators are proportional to the identity, since the channel $\Lambda_{\e}$ will result in the state to change which,  by the contractivity of the trace distance under quantum channels, implies that for any $\rho\sub{\s}$ and $\epsilon \in (0,1]$, there exists an $\alpha \in \mathds{N}$ such that $\frac{1}{2}\| \Lambda_{\e}^{(\alpha)}(\rho\sub{\s}) - \rho\sub{\s}^{\e}\|_1  \leqslant \epsilon$, where $\Lambda_{\e}^{(\alpha)}$ denotes $\alpha$ consecutive applications of  the quantum channel $\Lambda_{\e}$.

\end{document}